\documentclass[preprint,12pt,authoryear,times]{elsarticle}

\usepackage[T1]{fontenc}
\usepackage{graphicx}
\usepackage{amsmath}
\usepackage{amssymb}
\usepackage{booktabs}
\usepackage{array}
\usepackage{microtype}
\usepackage{hyperref}

\usepackage{enumitem}
\usepackage{multirow}
\usepackage{caption}
\usepackage{placeins}
\usepackage{float}
\usepackage{pdflscape}

\usepackage{geometry}
\journal{}
\biboptions{authoryear,round}
\hypersetup{colorlinks=true,citecolor=blue,linkcolor=blue,urlcolor=blue}
\makeatletter
\def\ps@pprintTitle{\let\@oddhead\@empty\let\@evenhead\@empty
\def\@oddfoot{\hfil\thepage\hfil}\let\@evenfoot\@oddfoot}
\makeatother

\begin{document}

\begin{frontmatter}

\title{Constraining Particle Stiffness in Asteroid Regolith from Early-Time High-Speed Penetrator Dynamics under Local Granular Variability}

\author[aff1]{Zichen Wang}
\ead{wang_zichen@buaa.edu.cn}
\author[aff2]{Yang Yu\corref{cor1}}
\ead{yu.yang@buaa.edu.cn}
\author[aff3]{Chenyang Huang}
\ead{chenyangh1314@gmail.com}
\author[aff4]{Zhijun Song}
\ead{songzhj9@mail.sysu.edu.cn}
\author[aff2]{Yutian Wu}
\ead{wuyutian@buaa.edu.cn}
\author[aff5]{Xiaojing Zhang}
\author[aff5]{Xuhui Zhang}
\ead{zhangxh0215@126.com}
\cortext[cor1]{Corresponding author: Yang Yu}
\address[aff1]{School of Future Aerospace Technology, Beihang University, Beijing 100191, China}
\address[aff2]{School of Aeronautic Science and Engineering, Beihang University, Beijing 100191, China}
\address[aff3]{Department of Aerospace Science and Technology, Politecnico di Milano, Via La Masa 34, 20156 Milano, Italy}
\address[aff4]{Planetary Environmental and Astrobiological Research Laboratory (PEARL), School of Atmospheric Sciences, Sun Yat-sen University, Zhuhai 519082, China}
\address[aff5]{China Academy of Aerospace System and Innovation, Beijing, China}

\begin{abstract}
Mechanical properties of rubble-pile asteroid regolith remain poorly constrained because contact responses depend on both material properties and local particle configuration. This study examines whether early-time high-speed penetrator dynamics can constrain particle stiffness, represented by the particle Young's modulus $E$, within a controlled discrete-element model. A bidirectionally coupled EDEM--Adams discrete-element--multibody model is used to screen six parameters: Young's modulus, Poisson's ratio, coefficient of restitution, static friction coefficient, rolling friction coefficient, and adhesion strength. Five Young's-modulus levels are then tested at the same 17 spatial locations in one settled granular bed, giving 85 simulations with location-matched comparisons across modulus levels. Within the 75--90~m~s$^{-1}$ speed range, varying Young's modulus yields the clearest and most systematic differences in probe deceleration, whereas the other parameters have weaker effects over the tested ranges. Validation by withholding all modulus cases at one location in turn indicates that the early response is more informative for distinguishing broad modulus ranges than adjacent modulus levels. Mesoscale analysis links the response differences to concentrated interface load sharing and spatially extended three-dimensional high-capacity contact paths as Young's modulus increases. The results provide a numerical basis for narrowing the particle Young's-modulus range from high-speed penetration responses despite local granular variability. Quantitative application to real asteroid regolith requires further experimental validation.
\end{abstract}


\end{frontmatter}

\section{Introduction}
\label{sec:introduction}

Asteroids preserve primitive material from the early Solar System and are therefore key targets for studies of planetary formation and evolution. Their surfaces, however, remain mechanically uncertain environments for sampling, anchoring, mobility, and other contact operations. Spacecraft observations have shown that many small bodies are not coherent monoliths but gravitational aggregates with boulder-rich and strongly heterogeneous surfaces. Hayabusa observations established the rubble-pile nature of Itokawa \citep{Fujiwara2006}, while OSIRIS-REx revealed the unexpectedly rough and boulder-dominated surface of Bennu \citep{Lauretta2019}. Under the extremely weak gravity of small bodies, interparticle contact and cohesive forces can become comparable to, or even exceed, particle weight \citep{Scheeres2010,Sanchez2014}. Local mechanical response is therefore governed strongly by particle arrangement and contact state and cannot be inferred reliably from terrestrial-soil experience alone.

This uncertainty is both a scientific issue and an engineering constraint. Previous missions have shown that local surface conditions directly affect landing, sampling, and contact dynamics. Hayabusa performed touchdown and sampling on Itokawa \citep{Yano2006}; Hayabusa2 encountered a loose and readily disturbed near-surface layer during sampling at Ryugu \citep{Morota2020}; and reconstruction of the multiple Philae touchdowns showed that local strength and layering can substantially modify lander dynamics \citep{Biele2015,Roll2016}. These mission experiences motivate methods for obtaining local mechanical information before or during surface operations.

Contact- and penetration-based characterization provide a direct route because the reaction of the target medium is encoded in the motion and load history of the probe. Philae contact dynamics were used to constrain comet-surface mechanical properties \citep{Biele2015}, and the OSIRIS-REx interaction with Bennu provided evidence for an extremely loose, near-zero-cohesion subsurface \citep{Walsh2022}. For operations that require active entry into the near surface, kinetic penetration offers an additional advantage: the required penetration energy can be imparted before contact, reducing the need to sustain a large quasi-static reaction force under microgravity \citep{Jiang2022}.

\begin{figure}[!htbp]
  \centering
  \includegraphics[width=0.78\linewidth]{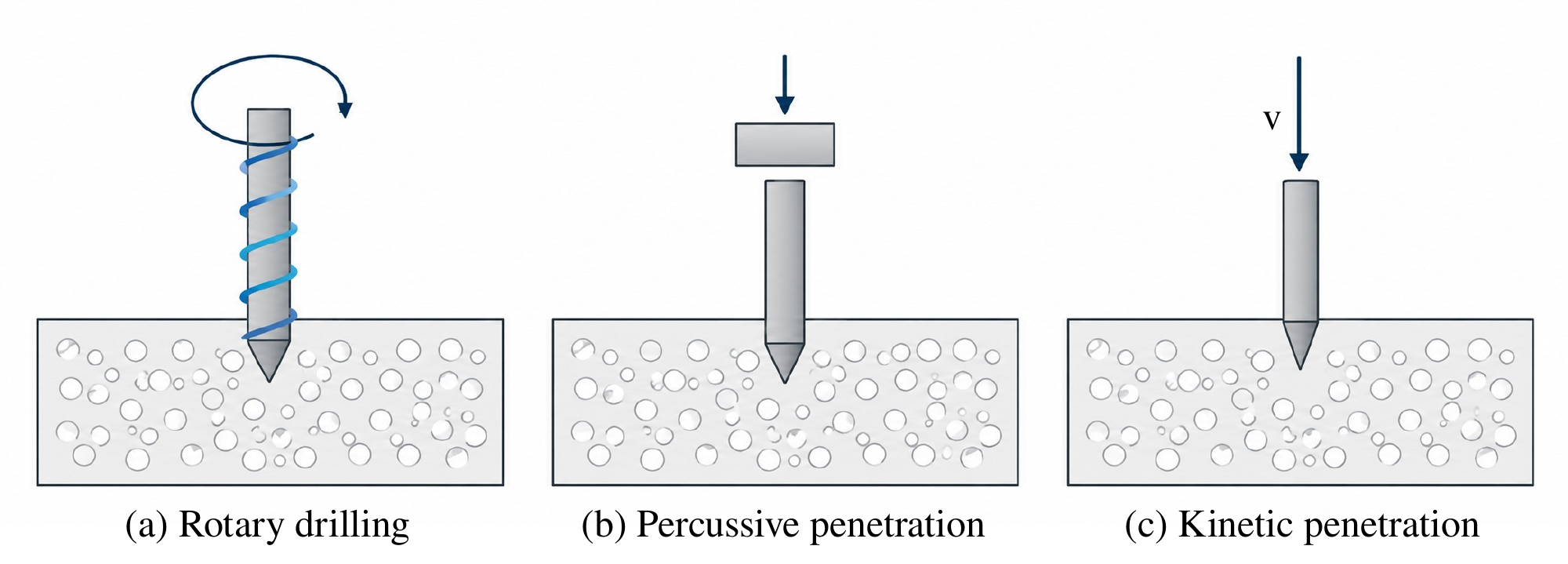}
  \caption{Schematic comparison of three penetration strategies for extraterrestrial regolith: (a) rotary drilling, (b) percussive penetration, and (c) kinetic penetration. This study focuses on high-speed kinetic penetration, in which the probe's impact kinetic energy drives entry into the regolith.}
  \label{fig:penetration_modes}
\end{figure}

Figure~\ref{fig:penetration_modes} compares three representative penetration strategies for extraterrestrial regolith: rotary drilling, percussive penetration, and kinetic penetration. Rotary drilling enables controlled excavation and sampling but generally requires continuous actuation and reaction force. Percussive penetration achieves incremental entry through repeated impacts. By contrast, a kinetic penetrator primarily uses translational kinetic energy acquired before contact. Deep Space~2 is a representative example of this concept \citep{Smrekar1999}. Although its target environment differs from that of rubble-pile asteroids, it demonstrates that penetration energy can be stored in the probe before impact. The associated cost is the large transient load that must be tolerated by the probe structure and onboard instrumentation.

Although kinetic-penetrator concepts have been proposed for planetary and small-body exploration, their use for remotely characterizing the mechanical properties of asteroid regolith remains largely unvalidated. In particular, few studies have examined the case in which a probe is launched from a distance, penetrates a rubble-pile surface, and uses its own transient motion to constrain local regolith properties. Two questions are central. First, the target is a discrete granular assembly, so local contact geometry and load-bearing paths vary with penetration location. A probe signal can therefore contain contributions from both material parameters and local configuration. DEM explicitly resolves particle contacts and is well suited to probe--regolith interaction problems \citep{Cundall1979,Maurel2018}. Because granular load transmission is strongly heterogeneous, a useful macroscopic signal must preserve a systematic material trend despite local configurational variability.

Second, the target parameter must be distinguishable from other contact parameters that also influence probe dynamics. In a soft-sphere DEM model, the response may depend simultaneously on particle Young's modulus, Poisson's ratio, restitution, friction, adhesion, and the initial packing state. If several parameters produce comparable changes in the early signal, reliable inference from the axial trajectory becomes difficult. It is therefore necessary to determine which parameters produce the clearest early response, whether that signature remains distinguishable across different local particle arrangements, and what particle-scale mechanism accompanies the observed macroscopic trend.

\begin{figure}[!htbp]
  \centering
  \includegraphics[width=0.96\linewidth]{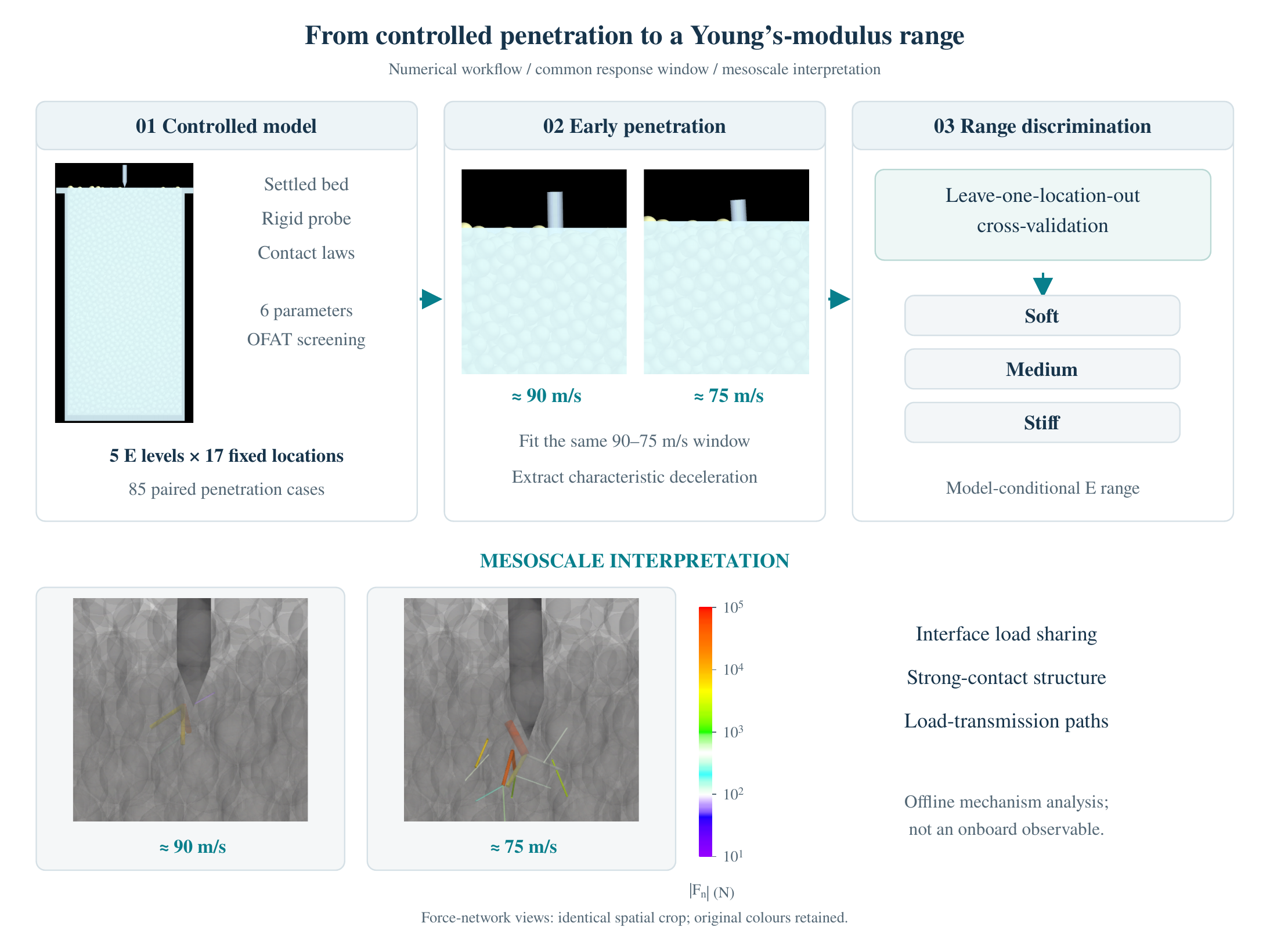}
  \caption{Controlled numerical workflow for constraining the particle Young's-modulus range from early penetrator dynamics. Five modulus levels are paired across 17 fixed locations; characteristic deceleration is extracted in the common 90--75~m~s$^{-1}$ window and evaluated by leave-one-location-out cross-validation. The lower panels show local contact-force networks near the two window endpoints and their role in offline mesoscale interpretation. The two contact views use the same spatial crop and the original logarithmic normal-force colour scale.}
  \label{fig:concept}
\end{figure}

This study investigates a probe-response-based numerical framework for constraining the Young's modulus of asteroid-regolith particles. Figure~\ref{fig:concept} presents the general stiffness-characterization workflow used here; in the present study, the calibrated stiffness variable is the particle Young's modulus $E$. A kinetic probe impacts the granular surface at high speed and records its axial velocity or acceleration. A characteristic deceleration is extracted from the early continuous-deceleration stage and compared with numerical or experimental calibration. Depending on calibration quality and response overlap, the result can be expressed as a supported modulus level, interval, or probability distribution. Mesoscale quantities, including coordination, strong-contact organization, interface load sharing, and internal load-transmission paths, are used only to interpret the physical origin of the macroscopic signal and the scatter caused by local configuration; they are not required onboard observables. A velocity of 100~m~s$^{-1}$ is adopted as a representative high-speed condition for evaluating method feasibility and is not proposed as a universal optimum for asteroid missions.

The study proceeds in three stages. First, six particle/contact parameters---particle Young's modulus, Poisson's ratio, restitution, static friction, rolling friction, and adhesion strength---are screened under otherwise identical high-speed penetration conditions to identify which parameter produces the clearest and most systematic separation in the early probe response; penetration-stage time-step independence is assessed separately in the numerical settings. Second, five representative $E$ levels are evaluated at 17 penetration locations, giving an 85-case ensemble that separates the systematic Young's-modulus trend from location-dependent scatter and quantifies the inference resolution supported by the probe response. The soft/medium/stiff grouping introduced later is treated as a coarse discretization supported by the present calibration, not as a universal resolution limit. Third, interface load sharing and weighted contact-network analyses are used to interpret how changes in $E$ reorganize contact forces and accompany changes in the early probe response. The long-term objective is to combine representative pre-mission numerical--experimental calibration with a limited amount of in situ probe data to constrain the Young's-modulus range of local asteroid-regolith particles; increasing calibration density alone is not assumed to eliminate overlap caused by granular variability.

\section{Numerical methodology and simulation setup}
\label{sec:methodology}

The target is represented as a controlled coarse-regolith analogue composed of decimeter-scale rubble particles. Because rubble-pile surfaces are intrinsically discrete, local contact geometry and load-transmission paths vary with position. Particle motion and contact forces are solved with the discrete element method (DEM), whereas penetrator motion under granular resistance is solved with multibody dynamics (MBD). The coupled DEM--MBD framework links the probe response to the underlying particle-scale contact process \citep{Cundall1979,Pang2013,Cheng2021}. The methodology comprises five elements: the penetrator and DEM--MBD coupling, the rubble-particle contact and dust-mediated adhesion model, preparation of the microgravity bed and the controlled simulation design, extraction and statistical analysis of the probe response, and mesoscale/contact-network analysis.

\subsection{Penetrator model and DEM--MBD coupling}
\label{sec:coupling}

\begin{figure}[!htbp]
  \centering
  \includegraphics[width=0.82\linewidth]{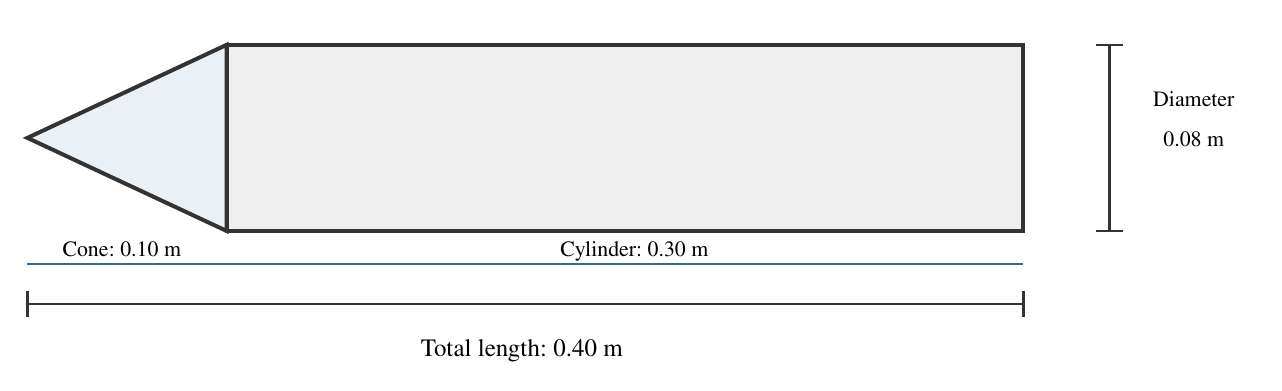}
  \caption{Steel cone--cylinder kinetic penetrator used in the simulations.}
  \label{fig:probe}
\end{figure}

The penetrator is a simplified steel cone--cylinder body based on representative dimensions used in kinetic regolith-penetration concepts \citep{Jiang2022}. As illustrated in Fig.~\ref{fig:probe}, the probe diameter is 0.08~m and the total length is 0.40~m, consisting of a 0.10~m conical nose and a 0.30~m cylindrical section. With a steel density of 7801~kg~m$^{-3}$, the model mass is approximately 13~kg. The probe Young's modulus is $2.07\times10^{11}$~Pa and Poisson's ratio is 0.29. The initial velocity is normal to the nominal bed surface. Previous high-speed penetrator and anchoring studies have used impact or penetration velocities of the same order: 90~m~s$^{-1}$ in an asteroid-anchor DEM study \citep{Chen2016}, $80\pm20$~m~s$^{-1}$ for the Mars-96 penetrator concept \citep{Surkov1998}, and about 180~m~s$^{-1}$ for the Deep Space~2 concept \citep{Smrekar1999}. These studies involve different bodies and target materials; they provide a precedent for the velocity scale, not a mission-specific optimum or a validation of the present particle response. Accordingly, 100~m~s$^{-1}$ is adopted as a literature-grounded, representative high-speed condition for testing the proposed response analysis. Impact-speed sensitivity is not evaluated. The decimeter-scale particles remain intact soft spheres throughout impact: no breakage criterion, fragment generation, or post-fracture contact law is implemented. This is a deliberate model assumption, not a claim that real asteroid blocks would survive impact at this speed.

\FloatBarrier

\begin{figure}[!htbp]
  \centering
  \includegraphics[width=0.98\linewidth]{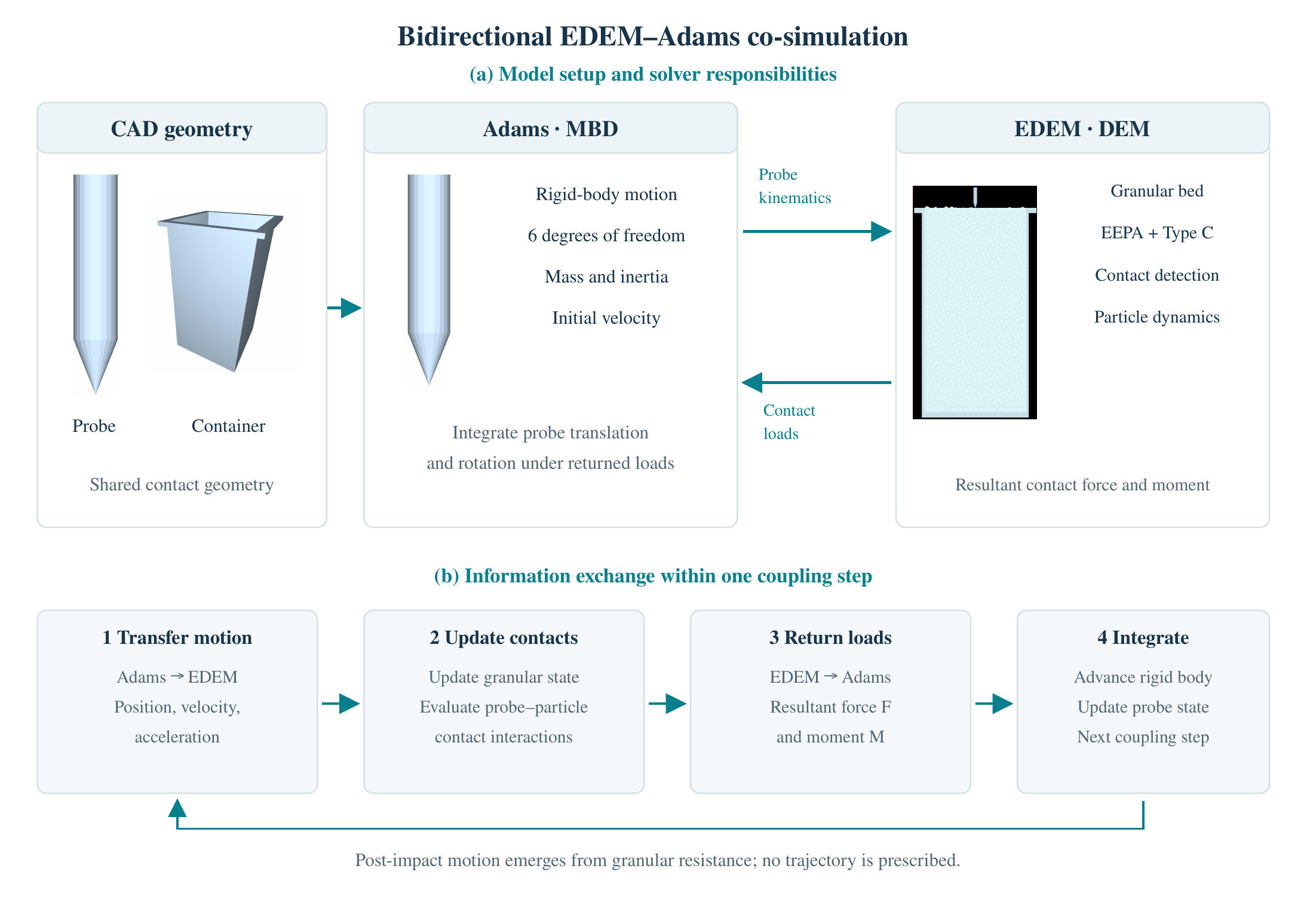}
  \caption{Bidirectional EDEM--Adams co-simulation: (a) geometric inputs and solver responsibilities; (b) motion transfer, particle/contact update, load return, and rigid-body integration within one coupling step. Probe motion after impact is determined by the returned contact loads rather than by a prescribed penetration trajectory. Geometry images illustrate model composition and are not shown to a common scale.}
  \label{fig:coupling}
\end{figure}

Particle-bed dynamics are solved in Altair EDEM 2024, while the penetrator is represented in MSC Adams 2020 as a six-degree-of-freedom rigid body. No post-impact trajectory is prescribed: the probe is initialized with the specified velocity and is then allowed to translate and rotate according to the forces and moments returned by the granular model. Although the full six-degree-of-freedom state is retained, the present inference feature uses only the global axial velocity component $v_z$, because the experiment is designed around nominally vertical penetration. The probe and container geometry are defined from CAD, the granular material and contact models are defined in EDEM, and the probe mass and initial state are defined in Adams. The overall co-simulation workflow and information exchange within one coupling step are shown in Fig.~\ref{fig:coupling}.

\FloatBarrier

At each coupling step, Adams transfers the current probe position, velocity, and acceleration to EDEM. EDEM updates the granular state, detects probe--particle contacts, evaluates the resultant force and moment on the probe, and returns these loads to Adams, which advances the rigid-body state. The resulting displacement and velocity histories therefore emerge from the instantaneous granular resistance rather than from prescribed post-impact kinematics. The production Adams command file spans 0.02~s with 100,000 requested dynamic steps, corresponding to a nominal requested interval of $2\times10^{-7}$~s. The recorded ACSI co-simulation settings use force absolute and relative tolerances of 0.01, interpolation-order settings of $-1$ for exchanged motion and force quantities, a linear weight of 0.9, and a transfer fraction of 0.5. These values are reported as implementation settings rather than interpreted as physical model parameters. A separate coupling-transfer-interval sensitivity sweep was not performed.

The baseline physical parameters are summarized in Table~\ref{tab:baseline}. The EDEM particle bed uses a prescribed gravitational acceleration of $1.3589\times10^{-5}$~m~s$^{-2}$, following the representative surface environment considered for 2016~HO3 (Kamo\textquotesingle oalewa) by \citet{LiScheeres2021}. The archived Adams model applies $1.0\times10^{-5}$~m~s$^{-2}$ to the probe. This solver-side difference is reported explicitly for reproducibility; over the 0.02-s penetration interval it changes the relative gravitational velocity and displacement by only $7.18\times10^{-8}$~m~s$^{-1}$ and $7.18\times10^{-10}$~m, respectively, and is negligible compared with the contact-driven response.

\begin{table}[H]
\centering
\caption{Baseline physical parameters of the probe--particle-bed model.}
\label{tab:baseline}
\small
\setlength{\tabcolsep}{3.5pt}
\begin{tabular}{llll}
\toprule
Category & Parameter & Symbol & Baseline value \\
\midrule
Particle & Diameter & $d$ & 0.10 m \\
 & Density & $\rho$ & 3400 kg m$^{-3}$ \\
 & Young's modulus & $E$ & $3.5\times10^{8}$ Pa \\
 & Poisson's ratio & $\nu$ & 0.25 \\
 & Shear modulus & $G=E/[2(1+\nu)]$ & $1.4\times10^{8}$ Pa \\
 & Particle--particle restitution & $e$ & 0.84 \\
 & Particle--particle static friction & $\mu$ & 0.80 \\
 & Particle--particle rolling friction & $\mu_r$ & 0.80 \\
 & Adhesion strength & $c$ & 10 Pa \\
\midrule
Environment & Gravitational acceleration & $g$ & $1.3589\times10^{-5}$ m s$^{-2}$ \\
\midrule
Probe & Mass & $m_p$ & $\approx13$ kg \\
 & Radius & $r_p$ & 0.04 m \\
 & Total length & $L_p$ & 0.40 m \\
 & Young's modulus & $E_p$ & $2.07\times10^{11}$ Pa \\
 & Poisson's ratio & $\nu_p$ & 0.29 \\
\midrule
Probe--particle & Restitution & $e_{pr}$ & 0.20 \\
 & Static friction & $\mu_{pr}$ & 0.60 \\
 & Rolling friction & $\mu_{r,pr}$ & 0.01 \\
\bottomrule
\end{tabular}
\end{table}

\subsection{Soft-sphere contact, equivalent adhesion, and rolling resistance}
\label{sec:contact}

The contact dynamics of the penetrating probe and the rubble bed are described with a soft-sphere discrete-element method. ``Soft sphere'' here does not imply that the real clasts are intrinsically soft; it means that a finite numerical overlap is permitted and used to represent local contact deformation. In contrast to an instantaneous binary-collision idealization, this formulation continuously resolves normal load, tangential slip, dissipation, and rolling resistance over a finite contact duration. It is therefore suitable for the dense, enduring multi-contact state and evolving load-bearing networks considered here. Soft-sphere DEM has been used extensively for the structural evolution of rubble-pile asteroids, regolith dynamics, and impact sampling \citep{Sanchez2011,Schwartz2012,Cheng2017}.

The decimeter-scale rubble is represented by spherical discrete elements with finite Young's modulus and Poisson's ratio. For contacting particles $i$ and $j$, the normal geometric overlap is
\begin{equation}
\delta_n=r_i+r_j-\left\|\mathbf{x}_i-\mathbf{x}_j\right\|,
\label{eq:overlap}
\end{equation}
and the non-adhesive elastic loading branch is chosen to retain the Hertz-type nonlinear relation \citep{Hertz1882}
\begin{equation}
F_{n,\mathrm{el}}=\frac{4}{3}E^*\sqrt{R^*}\,\delta_n^{3/2},
\label{eq:hertz_type}
\end{equation}
where
\begin{equation}
E^*=\left[\frac{1-\nu_i^2}{E_i}+\frac{1-\nu_j^2}{E_j}\right]^{-1},
\label{eq:effective_modulus}
\end{equation}
and
\begin{equation}
\frac{1}{R^*}=\frac{1}{r_i}+\frac{1}{r_j}.
\label{eq:effective_radius}
\end{equation}
Thus particle Young's modulus $E$ enters the normal elastic response through the pair effective modulus $E^*$. Normal energy dissipation is parameterized through the restitution coefficient. Tangential force is accumulated from relative tangential displacement through the EEPA tangential stiffness and limited by the model friction law. For contacts between particles made of the same isotropic elastic material, the Hertz--Mindlin incremental tangential-to-normal stiffness ratio is \citep{Mindlin1949}
\begin{equation}
\zeta_{\mathrm{tm}}=\frac{k_t}{k_n}=\frac{2(1-\nu)}{2-\nu}.
\label{eq:tangential_ratio}
\end{equation}
The baseline $\nu=0.25$ gives $\zeta_{\mathrm{tm}}=0.857143$. This is an elastic-consistency choice, not a fitted material constant: EDEM defines the multiplier as the tangential-to-virgin-loading stiffness ratio and notes that elastic values lie between $2/3$ and 1 and depend on $\nu$ \citep{AltairEEPA2025}. Because the production EEPA setting has zero contact plasticity and zero surface energy, the elastic ratio is used here. In the Poisson-ratio screen, it is updated with $\nu$ according to Eq.~\eqref{eq:tangential_ratio} and is treated as a derived quantity rather than an additional control variable. Rolling resistance is included to oppose relative particle rotation. No particle-fragmentation or clast-plasticity law is introduced.

Adhesion between resolved rubble particles is represented by a coarse-grained fine-dust bridge. Under vacuum and microgravity, micron- and submicron-scale grains occupying gaps between larger clasts can form a cohesive matrix through van der Waals interactions and transmit attraction to the surrounding clasts \citep{Scheeres2010,Sanchez2014}. Because the 0.10-m numerical particles do not explicitly resolve these fines, an equivalent bridge-area scale is defined as
\begin{equation}
A_{\mathrm{eff}}=4(\beta R^*)^2,
\label{eq:aeff}
\end{equation}
where $\beta$ is a dimensionless area factor. The present study fixes $\beta=1$ so that the adhesion scan uses one consistent coarse-grained area. The quantity $A_{\mathrm{eff}}$ is an operational area scale for unresolved dust action; it is neither the true Hertz contact-patch area nor a directly measured geometric area of the bed. The corresponding clast--clast adhesive-force magnitude is
\begin{equation}
F_{\mathrm{adh}}=cA_{\mathrm{eff}},
\label{eq:adhesion_force}
\end{equation}
where $c$ is a coarse-grained contact-scale adhesion strength. For two equal clasts of radius $r=0.050$~m, $R^*=0.025$~m and, with $\beta=1$,
\begin{equation}
A_{\mathrm{eff}}=2.5\times10^{-3}\ \mathrm{m^2}.
\label{eq:aeff_value}
\end{equation}
The parameter $c$ is not identified directly with the bulk or global cohesion of an asteroid surface. Because $F_{\mathrm{adh}}\propto c\beta^2$, uncertainty in the area factor and in the bridge stress is not separately identifiable in the present coarse-grained mapping; $\beta$ is therefore held fixed and $c$ is scanned across several orders of magnitude.

The particle--particle base contact is implemented with the Edinburgh Elasto-Plastic Adhesion (EEPA) model \citep{Thakur2014,AltairEEPA2025}. The distinctly plastic regime of that model is not used; instead, the elastic, constant-pull-off limit is adopted. The loading exponent is $n=1.5$, giving the Hertz-type $\delta_n^{3/2}$ loading branch documented for EEPA. The Contact Plasticity Ratio is $\lambda_p=0$, so virgin loading and unloading/reloading stiffnesses are equal, while Surface Energy is $\Delta\gamma=0$, disabling the load-dependent meso-contact adhesion term. The tensile exponent is held at $X=1.5$ as a fixed separation-branch implementation setting; its isolated influence is not investigated. The Tangential Stiffness Multiplier follows Eq.~\eqref{eq:tangential_ratio}, with a baseline value of 0.857143. Dust-mediated attraction is introduced only through the EEPA Constant Pull-Off Force,
\begin{equation}
f_0=-F_{\mathrm{adh}}=-cA_{\mathrm{eff}},
\label{eq:eepa_mapping}
\end{equation}
where the minus sign denotes the direction of the contact force; the EDEM Constant Pull-Off Force field is assigned the non-negative magnitude $|f_0|$. The baseline $c=10$~Pa therefore corresponds to a GUI input of 0.025~N. The physical variable reported throughout the paper is $c$; $f_0$ is the corresponding contact-level software parameter. This implementation creates no permanent interparticle bonds and does not use Bonding V2 or its bond-state/force/torque outputs.

Particle--particle rotation is represented with the EDEM Type~C rolling-friction model, which is chained to the EEPA base contact without altering the normal adhesion mapping. In the EDEM implementation \citep{Ai2011,AltairRolling2025}, the total rolling-resistance moment contains non-viscous and viscous components:
\begin{equation}
\begin{aligned}
\mathbf M_r&=\mathbf M_r^k+\mathbf M_r^d,\\
\mathbf M_r^k&=-k_r\boldsymbol\theta_r,\qquad
k_r=\lambda k_n\mu_r^2R_r^2,\\
\left|\mathbf M_r^k\right|&\leq \mu_rR_rF_n,\qquad
\mathbf M_r^d=-2\eta\sqrt{I_rk_r}\,\boldsymbol\omega_r.
\end{aligned}
\label{eq:typec}
\end{equation}
Here $\lambda$ is the coefficient of rolling stiffness, $\eta$ is the rolling viscous damping ratio, and $\mu_r$ is the rolling-friction coefficient of the relevant material pair. In this expression $F_n$ is the normal force used by the Type~C cap; the enabled ``Undamped Force Only'' option makes that cap use the undamped normal-force component. EDEM evaluates $k_n$, $R_r$, $I_r$, $\boldsymbol\theta_r$, and $\boldsymbol\omega_r$ from the base contact, particle properties, and instantaneous state. The official implementation applies the viscous term only while the non-viscous moment remains below its limiting value in Eq.~\eqref{eq:typec} \citep{AltairRolling2025}. The coefficient $\mu_r$ is not entered in the Type~C configuration dialog on the Physics page; it is assigned to the material pair under Creator--Bulk Materials--Interactions.

The production setting uses $\lambda=3$ and $\eta=0.3$, with ``Undamped Force Only'' and ``Tangential Rotation Only'' enabled. The first option bases rolling resistance on the undamped normal-force component; the second applies it only to relative rotation in the contact tangent plane, not to spin about the contact normal. The Polyhedral-to-Polyhedral Friction Model option is irrelevant to the spherical particles used here. Type~C supplies an effective rotational resistance for unresolved clast roughness, angularity, and local interlocking, but it does not explicitly resolve non-spherical geometry. Accordingly, $\lambda$ and $\eta$ are fixed mesoscale implementation parameters rather than directly measurable intrinsic rock constants.

Particle--geometry interactions (probe and container) use the EDEM default Hertz--Mindlin (no-slip) contact implementation together with Standard Rolling Friction; no EEPA pull-off term is applied to the artificial probe or container geometry. For the probe--particle pair, the restitution, static-friction, and rolling-friction coefficients are those listed in Table~\ref{tab:baseline}. Container material-pair coefficients are matched to the particle--particle baseline values where applicable. These implementation choices are summarized in Table~\ref{tab:contact_impl}.

\begin{table}[!htbp]
\centering
\caption{Contact-model implementation settings used in the production simulations.}
\label{tab:contact_impl}
\resizebox{\linewidth}{!}{%
\begin{tabular}{lll}
\toprule
Interaction & Model/parameter & Setting \\
\midrule
Particle--particle & EEPA loading exponent $n$ & 1.5 \\
 & Tensile exponent $X$ & 1.5 \\
 & Contact Plasticity Ratio $\lambda_p$ & 0 \\
 & Tangential Stiffness Multiplier & $2(1-\nu)/(2-\nu)$; baseline 0.857143 \\
 & Surface Energy $\Delta\gamma$ & 0 J m$^{-2}$ \\
 & Constant Pull-Off Force (GUI input) & $|f_0|=cA_{\mathrm{eff}}$; baseline 0.025 N \\
 & Rolling model & Type C \\
 & Particle--particle rolling friction $\mu_r$ & 0.80; set under Bulk Materials--Interactions \\
 & Rolling stiffness coefficient & 3 \\
 & Rolling viscous damping ratio & 0.3 \\
 & Type C options & Undamped Force Only; Tangential Rotation Only \\
\midrule
Particle--geometry & Normal/tangential model & Hertz--Mindlin (no slip), EDEM default \\
 & Rolling model & Standard Rolling Friction, EDEM default \\
\bottomrule
\end{tabular}%
}
\end{table}

The adhesion-strength range is treated as a physics-informed parameter scan rather than as a measured interval of single-contact asteroid adhesion. Observations and inversions for Bennu indicate very low macroscopic/local surface strength: impact-ejecta analysis gives a nominal local effective surface-strength upper bound of about 2~Pa \citep{Perry2022}, while structural-evolution modeling constrains surface cohesion compatible with widespread surface motion to $\lesssim0.78$~Pa \citep{ZhangYun2022}. These macroscopic values motivate inclusion of zero- and weak-adhesion cases but cannot be directly equated with the single-contact value of $c$. At the fine-grain matrix scale, \citet{Sanchez2014} obtained for random packings
\begin{equation}
\sigma_y=\frac{1.56\times10^{-4}}{\bar r_p},
\label{eq:sanchez_strength}
\end{equation}
where $\bar r_p$ is the mean fine-particle radius in metres and $\sigma_y$ is in Pa. For $\bar r_p=1~\mu$m, $\sigma_y\approx156$~Pa, corresponding to about 0.39~N over the present $A_{\mathrm{eff}}$; for $\bar r_p=0.1~\mu$m, $\sigma_y\approx1560$~Pa, corresponding to about 3.9~N. These estimates are conditional on the fine particles forming an effective load-bearing bridge over the adopted coarse-grained area. Accordingly, the screening uses
\begin{equation*}
c=0,\ 1,\ 10,\ 100,\ 1000,\ 1600\ \mathrm{Pa},
\end{equation*}
with corresponding EDEM Constant Pull-Off Force magnitudes (in N):
\begin{equation*}
|f_0|=0,\ 0.0025,\ 0.025,\ 0.25,\ 2.5,\ 4.
\end{equation*}
The range spans loose weak-adhesion conditions through a conditional upper-end case associated with submicron dust bridges. The 1600~Pa (approximately 4~N) endpoint is not interpreted as a measured or universal upper limit for Bennu or other rubble-pile asteroid surfaces.

\subsection{Bed preparation and controlled simulation design}
\label{sec:screening_setup}

\begin{figure}[!htbp]
  \centering
  \includegraphics[width=0.50\linewidth]{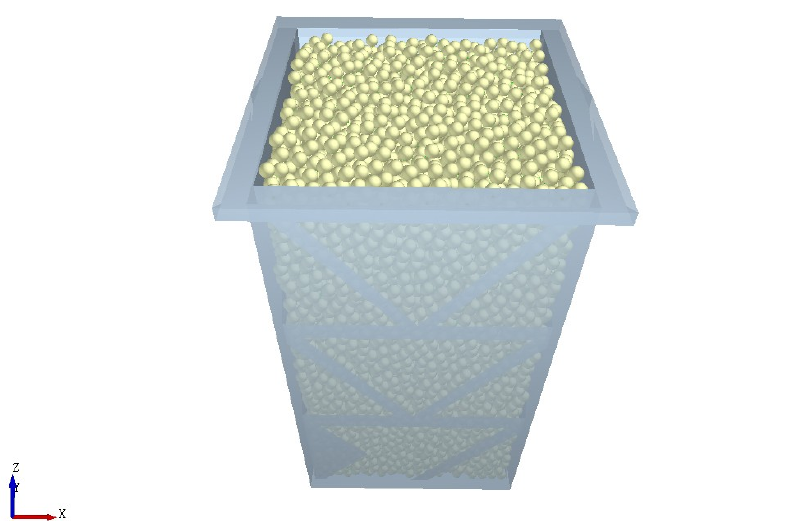}
  \caption{Granular bed after settling under the prescribed microgravity condition.}
  \label{fig:bed}
\end{figure}

A $2\times2\times4$~m container is used for the granular bed. With a 0.08~m probe diameter and 0.40~m probe length, the container width is 25 probe diameters and its depth is 10 probe lengths. The particle diameter is 0.10~m, giving $d/D_p=1.25$. These ratios document the scale of the numerical configuration; the model is not claimed to be a complete similarity-scaled reproduction of a particular asteroid mission. The domain also exceeds the relative dimensions used in previous intrusion studies to reduce direct boundary influence \citep{Zhang2022}. Monodisperse spherical particles are generated under the prescribed microgravity field and allowed to settle for 7500~s, producing a saved reference bed containing 16,700 particles (Fig.~\ref{fig:bed}). The monodisperse spherical representation is deliberate: it provides a controlled system in which material/contact-parameter effects and local configurational variability can be separated. It is not intended to imply that real asteroid regolith is monodisperse or spherical.

The same saved parent-bed state is reused deliberately as a control/blocking device. The initial bed is first settled to 7500~s. Whenever a material or contact parameter is changed, the same parent-bed state at 7000~s is reloaded and evolved for another 500~s under the new parameter setting before penetration begins. The archived 7000--7500~s \emph{baseline} history contains 26 saved states for 16,700 particles. Over these states the mean particle speed is $4.80\times10^{-7}$~m~s$^{-1}$ (maximum saved-state mean $5.10\times10^{-7}$~m~s$^{-1}$), the maximum single-particle speed is $1.84\times10^{-4}$~m~s$^{-1}$, and the total bed kinetic energy has a mean of $1.98\times10^{-8}$~J, a maximum of $6.17\times10^{-8}$~J, and a final value of $1.18\times10^{-8}$~J. These quantitative values apply to the baseline archive. For every changed-parameter case, the same 500-s re-equilibration protocol was applied and the final EDEM velocity field was inspected to confirm the absence of discernible bulk motion; a complete kinetic-energy history was not retained for every variant. The procedure allows the contact state to adjust while suppressing unrelated between-bed randomness. Reusing the same parent bed and, in the second stage, the same penetration coordinates therefore provides paired comparisons across $E$. The exact particle-generation random seed was not retained; reproducibility of the ensemble is instead anchored to the saved parent-bed state and the stated re-equilibration protocol.

The penetration stage uses Euler integration in EDEM with automatic time stepping disabled and a production DEM time step of $2\times10^{-7}$~s. For the reference setting, EDEM reports a Rayleigh time step of $8.4382\times10^{-4}$~s, so the fixed step is 0.0237\% of that value. Each penetration run spans 0.02~s. The target data-save interval for the probe-response histories is $2\times10^{-5}$~s; all post-processing uses the actual saved time stamps rather than assuming exact regular spacing. The EDEM GPU CUDA solver is selected; the production configuration lists 20 CPU cores and one GPU device. Impact-stage time-step independence is checked in the same baseline case by changing only the fixed step to $1\times10^{-7}$, $2\times10^{-7}$, and $4\times10^{-7}$~s. The three axial-velocity histories are indistinguishable at the present save and plotting resolution, so $2\times10^{-7}$~s is retained for production. This check applies to the 0.02-s high-speed penetration stage and is a response-level step-size check, not a formal convergence-order study over the full parameter space.

The first stage is an exploratory one-factor-at-a-time (OFAT) screening at a single common reference penetration configuration. Six particle/contact parameters are examined: particle Young's modulus $E$, Poisson's ratio $\nu$, particle--particle coefficient of restitution $e$, particle--particle static friction coefficient $\mu$, particle--particle rolling-friction coefficient $\mu_r$, and adhesion strength $c$. Within each family, only the target parameter is varied and all other parameters are held at their baseline values. Changing particle Young's modulus changes both particle--particle and particle--probe contact stiffness, since the particle contributes to the effective modulus of both material pairs. Changes in $e$, $\mu$, $\mu_r$, and $c$ refer only to the particle--particle material pair; probe--particle interaction coefficients remain at the separate values reported in Table~\ref{tab:baseline}. The tested values are selected from previous small-body surface-contact, regolith, penetration, and granular-dynamics studies and from the adhesion mapping above \citep{Maurel2018,Zhang2022,Cheng2021,Sanchez2014}; the baseline $E=3.5\times10^8$~Pa follows \citet{Maurel2018}. This first stage is used to identify a parameter family showing a clear early-response signature under one controlled reference configuration; it is not a global sensitivity analysis and does not establish unique identifiability when several parameters are simultaneously uncertain. The cases are listed in Table~\ref{tab:screening}.

\begin{table}[!htbp]
\centering
\caption{One-factor-at-a-time parameter-screening cases. Each row is an independent controlled virtual experiment: only the listed variable is changed, all other parameters use the Table~\ref{tab:baseline} baseline, and the six families are not combined as a Cartesian product.}
\label{tab:screening}
\begin{tabular}{>{\raggedright\arraybackslash}p{0.18\linewidth}>{\raggedright\arraybackslash}p{0.70\linewidth}}
\toprule
Control variable & Tested levels \\
\midrule
$E$ (Pa) & $5\times10^6$, $5\times10^7$, $1\times10^8$, $3.5\times10^8$, $7\times10^8$ \\
$\nu$ & 0.20, 0.25, 0.30, 0.35 \\
$e$ & 0.20, 0.50, 0.84, 1.00 \\
$\mu$ & 0.30, 0.60, 0.80, 1.00 \\
$\mu_r$ & 0.40, 0.60, 0.80 \\
$c$ (Pa) & 0, 1, 10, 100, 1000, 1600 \\
\bottomrule
\end{tabular}
\end{table}

The three rolling-friction runs belong to the single-reference OFAT screen, whereas the 17-position Young's-modulus ensemble is a separate set of penetration runs. Even when their nominal material parameters agree, numerical values from these two sets are not substituted for one another or treated as repeated measurements of the same trajectory. In the rolling-friction family, only the particle--particle material-pair coefficient $\mu_r$ under Bulk Materials--Interactions is changed. The Type~C coefficient of rolling stiffness $\lambda=3$, rolling viscous damping ratio $\eta=0.3$, and both enabled options remain fixed. Thus, $\mu_r$ is handled by the same OFAT rule as the other five material/contact parameters; $\lambda$ and $\eta$ are not control variables in this study.

Elastic constants are kept internally consistent with the EDEM material-property interface through
\begin{equation}
G=\frac{E}{2(1+\nu)}.
\label{eq:shear_modulus}
\end{equation}
The $E$ screen holds $\nu=0.25$, so that $G=0.4E$. The $\nu$ screen holds $E=3.5\times10^8$~Pa and updates both $G$ and the Tangential Stiffness Multiplier, as listed in Table~\ref{tab:nu_derived}. Neither $G$ nor the Tangential Stiffness Multiplier changes in the restitution, static-friction, rolling-friction, or adhesion-strength screens.

\begin{table}[!htbp]
\centering
\caption{Derived elastic parameters updated in the Poisson-ratio screen.}
\label{tab:nu_derived}
\begin{tabular}{ccc}
\toprule
$\nu$ & $G$ (Pa) & Tangential Stiffness Multiplier \\
\midrule
0.20 & $1.45833\times10^8$ & 0.888889 \\
0.25 & $1.40000\times10^8$ & 0.857143 \\
0.30 & $1.34615\times10^8$ & 0.823529 \\
0.35 & $1.29630\times10^8$ & 0.787879 \\
\bottomrule
\end{tabular}
\end{table}

The second stage tests whether the $E$ response identified in the exploratory screen persists when local granular configuration changes. Seventeen plan-view coordinates are used on the same parent-bed realization: $(0,0)$, $(\pm0.10,0)$, $(0,\pm0.10)$, $(\pm0.10,\pm0.10)$, $(\pm0.05,\pm0.15)$, and $(\pm0.15,\pm0.05)$~m. Each coordinate is simulated at all five $E$ levels, giving 85 paired cases. Reusing the same parent bed and coordinates is intentional: it controls global-bed variability so that differences across the five $E$ levels can be compared at corresponding local configurations. Re-equilibration after changing $E$ can adjust individual contacts, but the cases retain the same parent-bed realization and broadly corresponding local particle arrangements. The coordinates remain within the central region of the fixed container to sample local packing variations while maintaining comparable distances from the walls. Extending the sampling region toward the walls would introduce an additional change in boundary proximity. The 17 sites therefore sample \emph{local configurational variability within one controlled bed}; they are neither 17 independent global-bed realizations nor a spatial material-property map. This placement controls the relative wall geometry across the five $E$ levels at each site but does not establish that boundary effects vanish or that responses at neighboring sites are statistically independent.

\subsection{Probe-response feature extraction and statistical analysis}
\label{sec:response_analysis}

The following procedures are applied to the 85-case Young's-modulus ensemble and, for the reported rolling-friction comparison, to the three archived OFAT velocity histories. The coupled model retains full six-degree-of-freedom probe motion, but the inference feature is extracted from the global axial velocity $v_z(t)$. With positive $z$ upward, penetration corresponds to $v_z<0$ and the speed magnitude is $u(t)=|v_z(t)|$. Inspection of the common early continuous-deceleration regime motivated a fixed analysis window from $u=90$ to 75~m~s$^{-1}$. For each trajectory, threshold-crossing times are obtained by linear interpolation between adjacent saved samples; the 90~m~s$^{-1}$ crossing is set to relative time zero. The two interpolated points $(t_{90},90)$ and $(t_{75},75)$ are included alongside the original saved points strictly between them in the least-squares regression. The same rule is applied to the 85 cases and to the reported three-run rolling-friction comparison without case-specific retuning. Robustness of the 85-case classification is checked by repeating the complete extraction and LOSO analysis in the adjacent equal-width windows 92.5--77.5 and 87.5--72.5~m~s$^{-1}$; these windows test feature-placement sensitivity under the present probe scale and impact condition rather than impact-speed sensitivity.

Within this interval, speed is fitted by
\begin{equation}
u(t)=kt+b,\qquad a=-k,
\label{eq:achar}
\end{equation}
where $a>0$ is the characteristic deceleration used for inference. The coefficient of determination $R^2$ describes linear fit quality. The endpoint-based average deceleration $15/(t_{75}-t_{90})$ is used as a consistency check on the fitted slope; because it is computed from the same response window, it is not an independent validation statistic. For each $E$ level, the 17 site values are summarized by the sample mean and sample standard deviation. The five group means are also fitted in log--log coordinates by the descriptive power law $\bar a=CE^p$, where $p$ denotes the empirical response exponent and is kept distinct from the EEPA loading exponent $n$. This fit is used only to summarize the present calibration range.

To separate the systematic $E$ term from the paired location effect within the observed 85-case ensemble, an additive blocked model is applied to $\ln a$ \citep{Montgomery2017}:
\begin{equation}
\ln a_{ij}=\mu_0+\alpha_i+\gamma_j+\varepsilon_{ij},
\label{eq:blockmodel}
\end{equation}
where $\alpha_i$ denotes the $i$th Young's-modulus level and $\gamma_j$ the $j$th penetration-location block. Because there is one observation in each $E\times$location cell, $\varepsilon_{ij}$ cannot be separated into pure measurement error and interaction; it contains $E\times$location non-additivity together with other unmodeled variation. The corresponding quantities are therefore reported as an \emph{observed sum-of-squares partition} of this 85-case ensemble, not as population variance components or inferential ANOVA effect fractions.

Predictive discrimination is assessed by leave-one-location-out (LOSO) cross-validation \citep{Stone1974}. In each fold, all five $E$ responses from one coordinate are removed together, so that exact coordinate is absent from training. For a discrete class $q$, the training responses define a one-dimensional Gaussian likelihood
\begin{equation}
p(a\mid q)=\frac{1}{\sqrt{2\pi s_q^2}}\exp\left[-\frac{(a-\bar a_q)^2}{2s_q^2}\right],
\label{eq:gaussian_likelihood}
\end{equation}
where $\bar a_q$ and $s_q$ are estimated from the training fold. With equal class priors, the held-out response is assigned to the class with the largest likelihood \citep{Hastie2009}. Performance is summarized by accuracy, balanced accuracy, and macro-averaged F1. The same procedure is also evaluated after exploratory grouping into three broad intervals: soft ($5\times10^6$~Pa), medium ($5\times10^7$ and $1\times10^8$~Pa), and stiff ($3.5\times10^8$ and $7\times10^8$~Pa). Because the contact-overlap check later identifies $5\times10^6$~Pa as a large-deformation numerical end member, two post-hoc endpoint-sensitivity diagnostics are additionally reported: four-level LOSO after excluding that endpoint, and two-class medium-versus-stiff LOSO on the remaining 68 responses. The power-law fit is likewise repeated for the four remaining group means. These diagnostics test whether the principal ordering is created solely by the soft endpoint; they do not replace the original five-level analysis. The class labels group the sampled calibration levels. Their minimum and maximum values describe the sampled span of each class, rather than a validated continuous prediction interval or an intrinsic resolution limit. Because neighboring sites can share particles and surrounding structure within the same bed, LOSO is explicitly interpreted as a \emph{within-bed unseen-location transfer test}, not as independent-bed validation. It is also a closed-set test of the calibrated discrete levels/classes; continuous interpolation between unseen $E$ values is not validated here.

The numerical feature extraction uses ideal simulated $v_z(t)$ histories. Sensor noise, finite bandwidth, sampling jitter, quantization, filtering, and estimator latency are not introduced. The classification results therefore quantify model-level discrimination under the present numerical conditions rather than flight-instrument performance.

\subsection{Local-configuration and mesoscale analysis}
\label{sec:meso_methods}

Two levels of post-processing are used to distinguish bulk local packing from the rapidly evolving load-bearing structure. First, the initial structure around each penetration site is characterized by ordinary coordination number $Z=2N_c/N_p$, where $N_c$ is the number of internal particle--particle contact pairs and $N_p$ is the number of particles in the sampling domain \citep{Aste2006}. No rattler-removal correction is applied to this ordinary coordination number. Local porosity is estimated by Monte Carlo volume sampling in cylindrical regions centered on each penetration coordinate; 15,000 random points are used per ROI with post-processing seed 20260804. Points falling inside the physical particle spheres are counted as solid and all others as void. Three initial-structure ROIs are available ($R=2d,3d,4d$ with $L=4d,6d,8d$); the principal values reported in the multisite comparison use $R=4d$, $L=8d$. The first-contact gap and local surface-height range are also recorded.

Second, contact-force organization is analyzed over the 90--75~m~s$^{-1}$ response interval for one retained common reference configuration at each of the five $E$ levels. These five mechanism cases are illustrative case histories associated with the response study, not averages over all 17 locations; no ensemble-wide mechanism statistic is inferred from them. Dynamic cylindrical ROIs are referenced to the instantaneous probe axis/tip so that the same near-probe volume is compared as penetration proceeds. An internal particle--particle edge is admitted only when both endpoint particle centers lie inside the selected ROI. For each admitted contact, $F_n$ denotes the compressive normal-force magnitude used in the load-bearing analysis; tensile adhesive contacts remain active in the DEM dynamics but are not treated as high-capacity compressive transmission edges. A contact is classified as a \emph{strong contact} when $F_n$ exceeds the frame-wise mean $\langle F_n\rangle$ of the compressive contacts in that ROI. Mean-force separation of strong and weak subnetworks has precedent in granular mechanics \citep{Radjai1998,Peters2005}. The term ``strong contact'' is used rather than assigning every such edge to a force chain, because force-chain definitions generally require additional geometric continuity and direction criteria \citep{Peters2005}. Three ROIs are compared: $R=2d,3d,4d$, with corresponding axial lengths $L=4d,6d,8d$. The retained mechanism series contain 240, 45, 60, 47, and 36 frames from the lowest to highest $E$; their median output interval is $2\times10^{-5}$~s except for the $5\times10^7$~Pa mechanism export, for which it is $4\times10^{-5}$~s. Actual time stamps are used in every case.

The instantaneous probe--regolith interface is characterized by the number of directly contacting particles $N_{\mathrm{probe}}$ and an effective load-bearing count based on participation number \citep{Makse2000,Deng2017}. Let $\hat{\mathbf e}_p$ point along the penetration direction and let $\mathbf F_i^{(p)}$ be the total force exerted by directly contacting particle $i$ on the probe. The positive resistance contribution is
\begin{equation}
F_i=\max\!\left(0,-\mathbf F_i^{(p)}\cdot\hat{\mathbf e}_p\right),
\label{eq:interface_force}
\end{equation}
and
\begin{equation}
N_{\mathrm{eff}}=\frac{\left(\sum_iF_i\right)^2}{\sum_iF_i^2}=\frac{1}{\sum_i p_i^2},\qquad p_i=\frac{F_i}{\sum_iF_i}.
\label{eq:neff}
\end{equation}
The largest-one and largest-three load fractions are calculated from the same positive-resistance contributions; if fewer than three positive contributors are present, the top-three fraction is the sum of the available contributors.

For spatial organization, strong contacts form a weighted graph in which particles are nodes and compressive strong contacts are edges with weight $F_{n,e}$ \citep{Bassett2015}. The analysis records strong-contact edge count, connected-component count, and the axial/radial extents of probe-related components. The force-weighted axial-orientation descriptor is
\begin{equation}
A_F=\frac{\sum_{e\in\mathcal E_s}F_{n,e}\left(\mathbf n_e\cdot\hat{\mathbf e}_p\right)^2}{\sum_{e\in\mathcal E_s}F_{n,e}},
\label{eq:af}
\end{equation}
where $\mathcal E_s$ is the strong-contact edge set and $\mathbf n_e$ is the unit contact normal. Larger $A_F$ denotes stronger average alignment with the penetration axis. Morphology descriptors are sampled at the 16 common speed states $|v_z|=90,89,\ldots,75$~m~s$^{-1}$ and summarized by medians and interquartile ranges.

Finally, an operational widest-path metric quantifies how far a finite force bottleneck remains connected from the probe into the bed. The construction is related to graph-based/network-flow descriptions of granular load transmission \citep{Tordesillas2015,Patino2023}, but the specific metric is defined here for the present analysis. A virtual probe-root node $s$ is added to the particle network. Each directly contacting source particle $i$ is connected to $s$ by an edge of capacity equal to its positive axial \emph{normal} resistance contribution from the probe contact, while each internal particle--particle edge has capacity equal to its compressive normal force $F_{n,e}$. For any candidate root-to-particle path $P$, the path capacity and widest-path capacity are
\begin{equation}
B(P)=\min_{e\in P}w_e,\qquad C_j=\max_{P:s\rightarrow j}B(P),\qquad \rho_j=\frac{C_j}{F_{\mathrm{probe}}},
\label{eq:widest}
\end{equation}
where $F_{\mathrm{probe}}$ is the instantaneous positive axial resistance of the probe. The widest-path graph uses all compressive internal contacts satisfying the $R=4d$, $L=8d$ ROI rule, rather than only the strong-contact subset. A particle is reachable when $\rho_j\ge1\%$. The root edge is included in the bottleneck calculation; therefore a directly contacted source particle can be counted as reachable even when no qualifying internal particle--particle edge extends beyond it. The reported metrics are the fraction of analyzed frames with at least one reachable node, $P_{\mathrm{reach}}$, the mean reachable-particle count $N_{\mathrm{reach}}$, and the axial/radial transmission extents $L_{\mathrm{ax}}$ and $L_{\mathrm{rad}}$ relative to the probe tip and axis. $N_{\mathrm{reach}}$ and both extents are arithmetic means over every retained frame in the 90--75~m~s$^{-1}$ interval; frames without a reachable node contribute zero. The transmission aspect ratio in Fig.~\ref{fig:transmission_metrics}(d) is the ratio of these two all-frame means, $\overline{L}_{\mathrm{ax}}/\overline{L}_{\mathrm{rad}}$, rather than the mean of frame-wise ratios; this definition remains finite when some frames have no qualifying path. The 1\% level is an operational reference threshold, and robustness is checked over 0.01--5\%. The widest-path construction is a graph descriptor of contact-force connectivity, not an independent force-balance or dynamic-equilibrium solution.

\section{Results and discussion}
\label{sec:results}

The results are presented from the macroscopic probe response to the particle-scale load-transfer mechanism. Section~\ref{sec:sensitivity} compares the six controlled parameter families, Section~\ref{sec:robustness} evaluates the Young's-modulus response across the 17 local penetration configurations, and Section~\ref{sec:mesoscale} uses the contact-network metrics defined in Section~\ref{sec:meso_methods} to interpret the macroscopic trend.

\subsection{Parameter sensitivity and the Young's-modulus-dominated early signal}
\label{sec:screening_results}
\label{sec:sensitivity}

\begin{figure}[!t]
  \centering
  \includegraphics[width=0.98\linewidth]{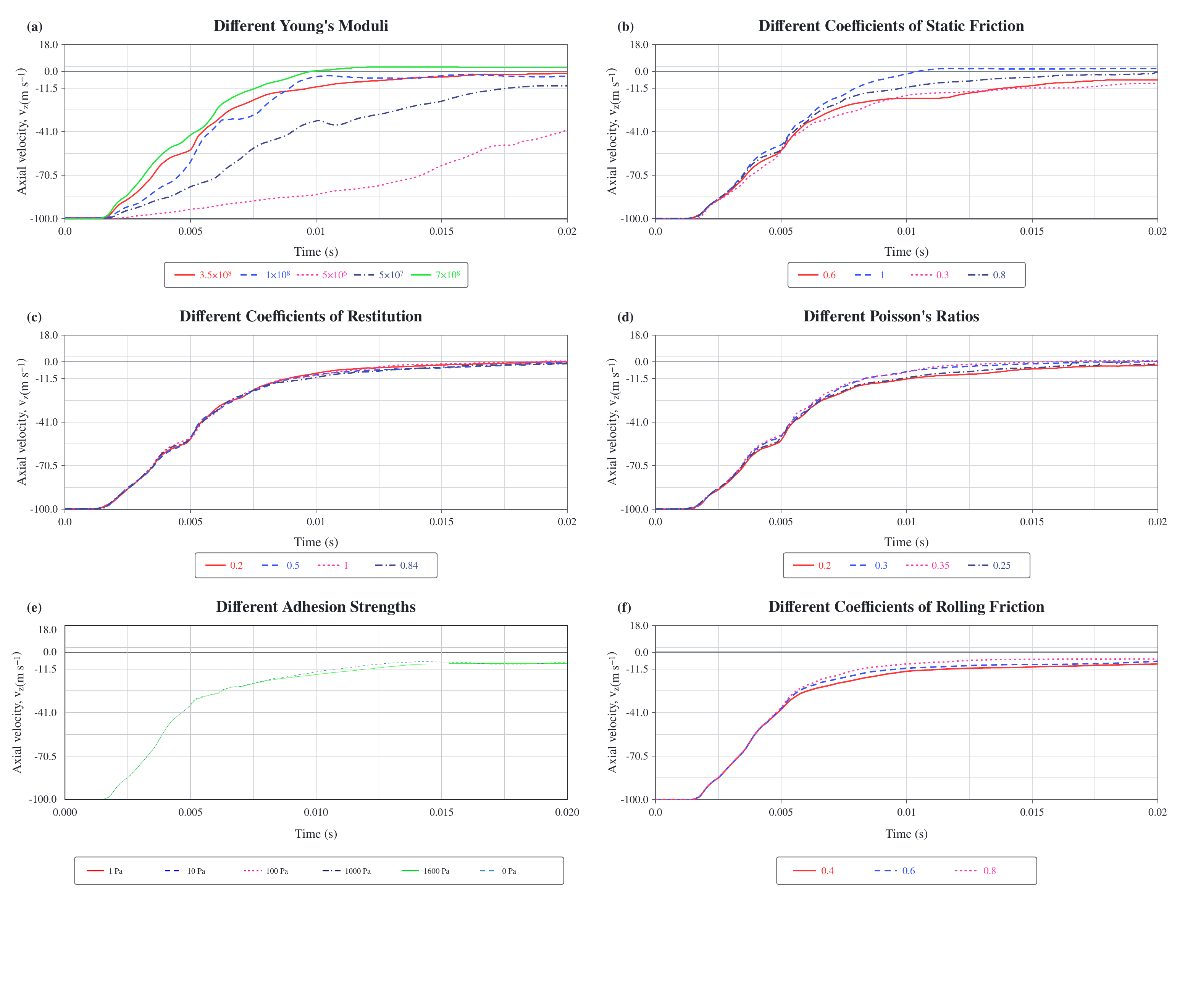}
  \caption{Axial velocity histories from the one-factor-at-a-time screening simulations: (a) particle Young's modulus, (b) particle--particle static friction coefficient, (c) particle--particle coefficient of restitution, (d) Poisson's ratio, (e) adhesion strength, and (f) particle--particle rolling friction coefficient. Panel (f) is replotted from the archived velocity samples using the same axes as the other panels.}
  \label{fig:screening}
\end{figure}

In the single-reference-configuration OFAT screen, varying the adhesion strength $c$ from 0 to 1600~Pa produces nearly overlapping probe-velocity histories over the early high-speed stage, including the 90--75~m~s$^{-1}$ interval used for the characteristic deceleration. Small differences become more visible only at later stages of penetration. Within this physics-informed range and reference configuration, dust-mediated adhesion therefore acts as a secondary correction rather than a governing factor of the early response. This result does not imply that cohesion is negligible in other asteroid-regolith conditions. Variations in Poisson's ratio over 0.20--0.35 likewise cause relatively small changes. Changing restitution from 0.20 to 1.00 does not produce a stable and pronounced separation of the early trajectories. Static friction has a more visible influence, but its effect becomes clearer only after substantial deceleration and remains weaker than the Young's-modulus effect in the same early interval. For particle--particle rolling-friction coefficients $\mu_r=0.4$, 0.6, and 0.8, the early velocity histories also nearly overlap (Fig.~\ref{fig:screening}(f)). Later separation shows that this early-stage result does not extend to the entire penetration history.

The Young's-modulus cases show a distinct response. Increasing $E$ from $5\times10^6$ to $7\times10^8$~Pa systematically accelerates the decay of probe speed, with separation already visible early in penetration. Among the six parameter families in this exploratory reference-configuration screen, $E$ therefore produces the clearest and most systematic early-response separation. This observation motivates the paired 17-location second stage, which tests whether the ordering persists under local configurational variability. This ordering depends on the tested ranges and on which interactions are varied: changing $E$ affects both particle--particle and probe--particle contact response, whereas the restitution, friction, and adhesion scans change the particle--particle pair while retaining the probe-interface coefficients. The comparison therefore supports a conditional screening result; it does not establish a global sensitivity ranking or unique identifiability when several parameters are uncertain.

\FloatBarrier

\subsection{Young's-modulus response under local configurational variability}
\label{sec:robustness}

\begin{figure}[!htbp]
  \centering
  \includegraphics[width=0.78\linewidth]{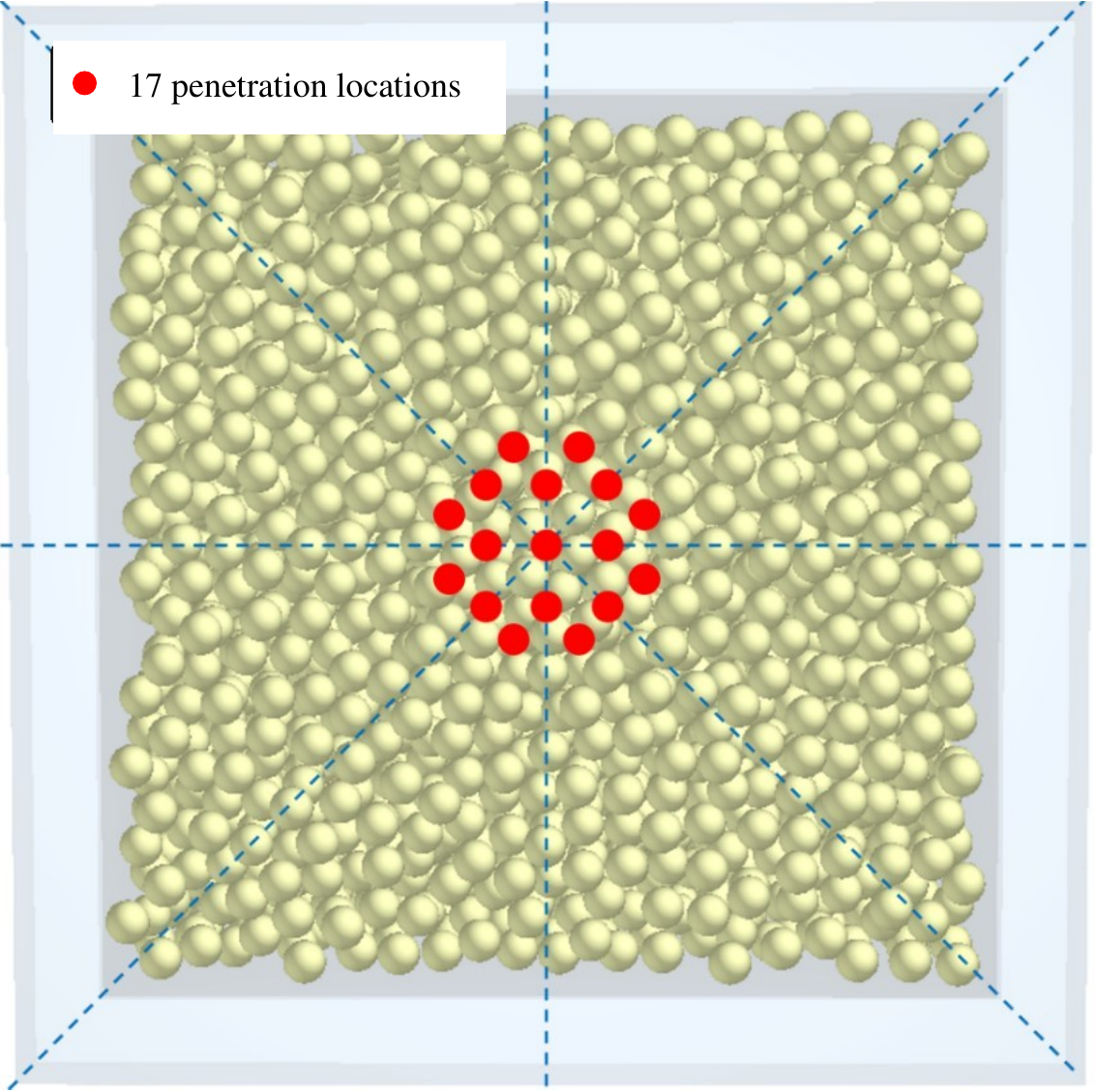}
  \caption{Plan-view distribution of the 17 penetration locations used to sample different local particle arrangements.}
  \label{fig:locations}
\end{figure}

Following the sequential design described in Section~\ref{sec:screening_setup}, the five Young's-modulus levels are evaluated at 17 penetration sites within the same settled parent-bed realization. Reuse of the parent bed and the same 17 coordinates is a deliberate control/blocking strategy: it suppresses unrelated between-bed variability and permits paired comparison of the five $E$ levels at corresponding local configurations. Figure~\ref{fig:locations} shows the plan-view locations. The sites sample different local surface and subsurface arrangements; they are not intended to represent a spatial material-property map or 17 statistically independent global bed realizations.

\FloatBarrier

\begin{landscape}
\begin{figure}[p]
  \centering
  \includegraphics[width=0.97\linewidth]{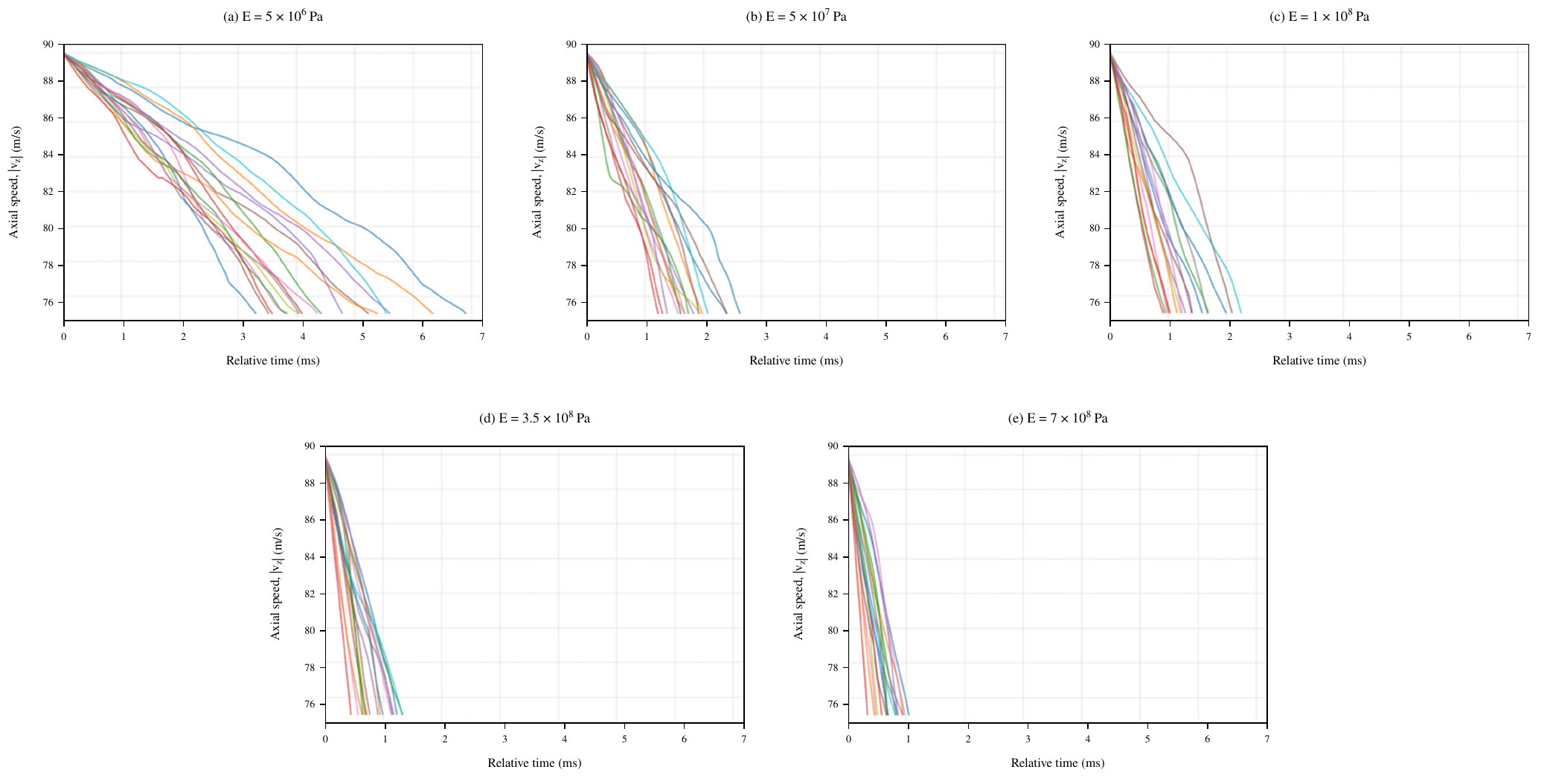}
  \caption{Early velocity histories for 17 penetration locations at five particle Young's moduli, aligned at $|v_z|=90$~m~s$^{-1}$: (a) $5\times10^6$~Pa, (b) $5\times10^7$~Pa, (c) $1\times10^8$~Pa, (d) $3.5\times10^8$~Pa, and (e) $7\times10^8$~Pa. Each curve corresponds to one penetration location; colours consistently identify the paired locations across all five panels.}
  \label{fig:aligned_velocity}
\end{figure}
\end{landscape}

The raw velocity histories show location-dependent variation in the onset of pronounced deceleration, primarily because of local surface height and first-contact geometry. Using the fixed 90--75~m~s$^{-1}$ extraction procedure in Section~\ref{sec:response_analysis}, the aligned early trajectories retain within-$E$ scatter but exhibit a common near-linear continuous-deceleration interval. Across the five Young's-modulus groups, the mean $R^2$ values lie between 0.981 and 0.990, and the overall mean over all 85 trajectories is approximately 0.986. Figure~\ref{fig:aligned_velocity} shows the 17 aligned trajectories for each $E$ level.

\FloatBarrier

Table~\ref{tab:multisite} summarizes the fitted characteristic-deceleration statistics. The mean increases from $3.35\times10^3$~m~s$^{-2}$ at $E=5\times10^6$~Pa to $2.23\times10^4$~m~s$^{-2}$ at $7\times10^8$~Pa. At the same time, the within-group standard deviation increases from $0.74\times10^3$ to $8.09\times10^3$~m~s$^{-2}$, showing that the absolute spread associated with local configuration also increases with $E$. The response distributions of several higher-$E$ groups therefore overlap even though their means remain ordered. Because the 17 sites are paired locations within one parent bed rather than independent bed realizations, Table~\ref{tab:multisite} reports descriptive mean and standard deviation without population-level confidence or prediction intervals.

\begin{table}[!htbp]
\centering
\caption{Descriptive multisite statistics of the fitted 90--75~m~s$^{-1}$ characteristic deceleration ($n=17$ locations per $E$ level within one parent bed).}
\label{tab:multisite}
\begin{tabular}{rrrr}
\toprule
$E$ (Pa) & Mean $a$ (m s$^{-2}$) & SD (m s$^{-2}$) & Mean $R^2$ \\
\midrule
$5\times10^6$ & 3350 & 738 & 0.989 \\
$5\times10^7$ & 8230 & 1756 & 0.981 \\
$1\times10^8$ & 11533 & 3432 & 0.990 \\
$3.5\times10^8$ & 17988 & 6494 & 0.986 \\
$7\times10^8$ & 22302 & 8086 & 0.986 \\
\bottomrule
\end{tabular}
\end{table}

\begin{landscape}
\begin{figure}[p]
  \centering
  \includegraphics[width=0.99\linewidth]{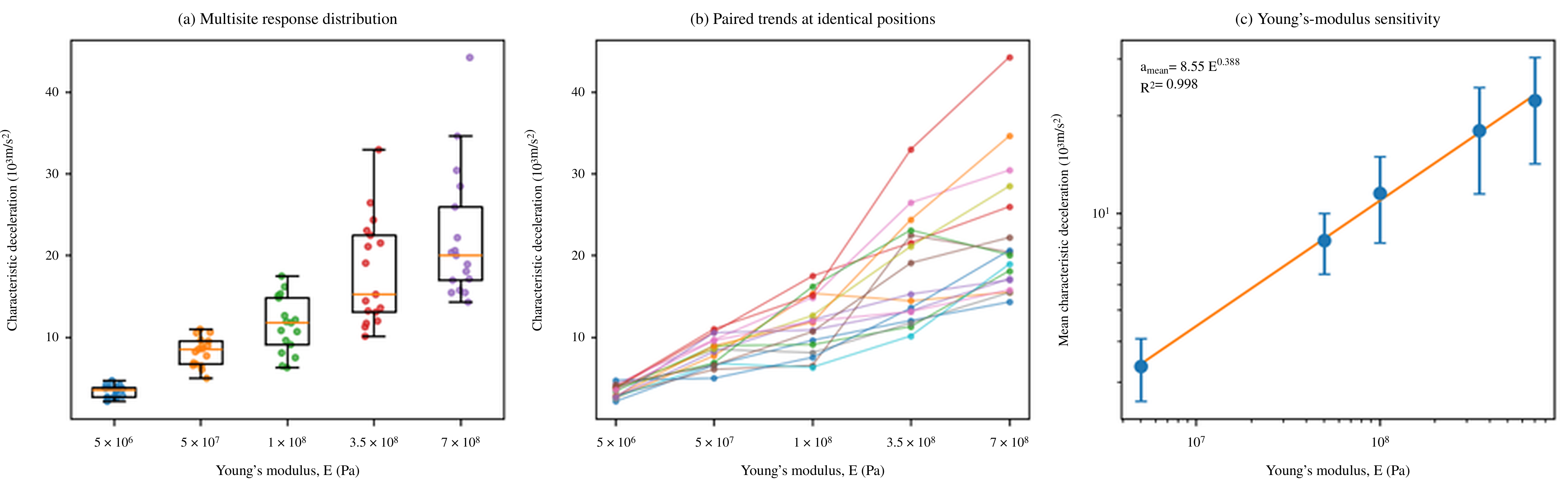}
  \caption{Statistical Young's-modulus signature from the 85 simulations: (a) multisite response distribution at each Young's modulus, (b) paired trends for the same 17 local configurations, and (c) empirical power-law fit to the location-averaged response. Error bars in (c) denote one sample standard deviation.}
  \label{fig:stiffness_stats}
\end{figure}
\end{landscape}

The paired-location comparison in Fig.~\ref{fig:stiffness_stats}(b) is particularly informative because the same spatial coordinates and parent-bed realization are reused at all five $E$ levels. Twelve of the 17 locations are strictly monotonic over all five levels, and every location gives a larger deceleration at $7\times10^8$~Pa than at $5\times10^6$~Pa. A log--log least-squares fit to the five group means gives an empirical response exponent $p=0.388$ with $R^2=0.998$. Excluding the large-deformation $5\times10^6$~Pa endpoint gives $p=0.374$ and $R^2=0.993$ for the remaining four means. Thus the positive mean trend is retained after excluding the soft endpoint. Both fits describe the group means within the present paired ensemble; they do not quantify population-level uncertainty or define universal constitutive relations.

\FloatBarrier

The observed sum-of-squares partition of $\ln a$ assigns 84.30\% to the Young's-modulus term, 8.44\% to the penetration-location block, and 7.26\% to the residual. With one observation per $E\times$location cell, the residual necessarily includes modulus--configuration non-additivity together with other unmodeled variation; it is not a pure experimental-error term. The percentages are therefore descriptive partitions of this paired 85-case ensemble, not population variance components or universal physical contribution fractions.

The initial-structure diagnostics show that bulk-average packing varies little across the 17 sites, whereas free-surface descriptors vary much more strongly. In the $R=4d$, $L=8d$ domain, the mean porosity is 0.4828 with a coefficient of variation of 1.15\%, while the mean coordination number is 6.261 with a coefficient of variation of 0.95\%. By contrast, the coefficient of variation of the first-contact gap is approximately 44.3\%, and that of the local surface-height range is approximately 41.1\%. This contrast is consistent with a role for near-surface and local-contact geometry, but by itself does not establish which descriptor causally dominates the deceleration scatter. Aligning the velocity histories removes much of the timing offset associated with first contact but does not eliminate the subsequent configurational scatter.

\begin{landscape}
\begin{figure}[p]
  \centering
  \includegraphics[width=0.98\linewidth]{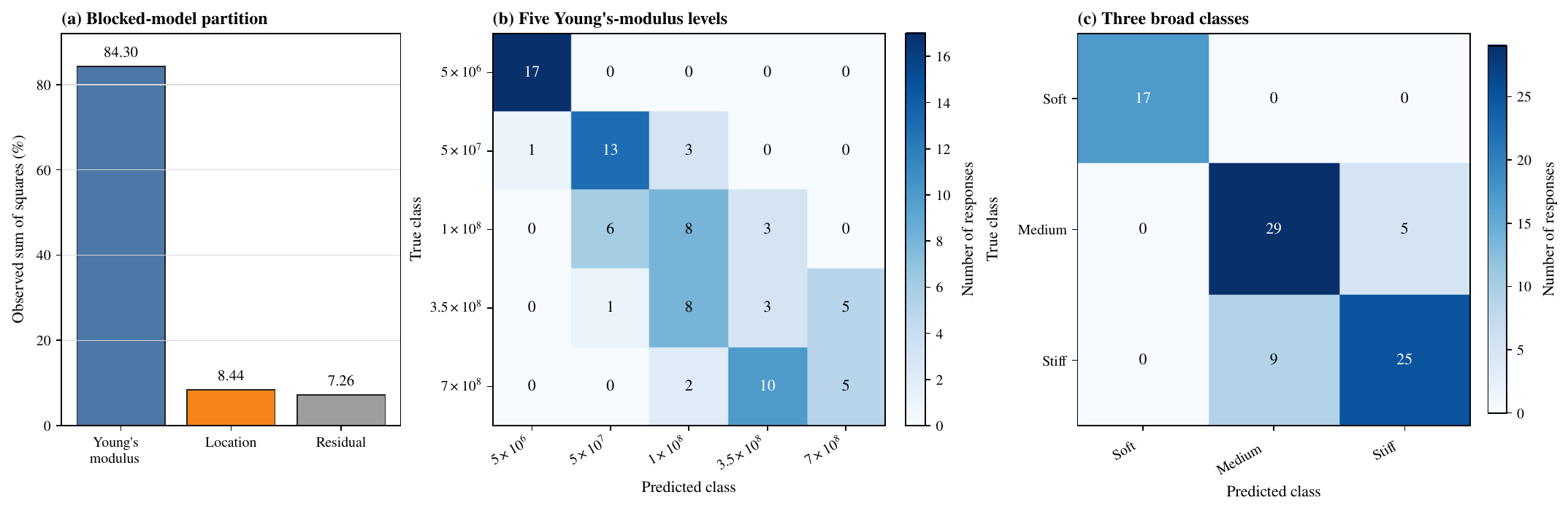}
  \caption{Robustness and Young's-modulus identification results for the paired 85-case ensemble: (a) observed sum-of-squares partition of $\ln a$ for the additive blocked model, (b) leave-one-location-out confusion matrix for five discrete $E$ levels, and (c) confusion matrix after exploratory grouping into three broad modulus classes.}
  \label{fig:classification}
\end{figure}
\end{landscape}

The LOSO classification results quantify the within-bed closed-set discrimination supported by the present sparse calibration. The five-level classifier correctly identifies 46 of 85 held-out responses: accuracy and balanced accuracy are both 54.12\%, and macro-F1 is 52.95\%. The class-wise correct-identification rates are 17/17 (100\%), 13/17 (76.5\%), 8/17 (47.1\%), 3/17 (17.6\%), and 5/17 (29.4\%) from the lowest to the highest $E$ level, showing marked overlap in the higher-modulus range. After exploratory grouping into the three broader modulus classes defined in Section~\ref{sec:response_analysis}, 71 of 85 responses are classified correctly, giving 83.53\% accuracy, 86.27\% balanced accuracy, and 86.23\% macro-F1. Figure~\ref{fig:classification} summarizes the observed sum-of-squares partition and both confusion matrices.

The adjacent-window check gives the same practical interpretation (Table~\ref{tab:window_sensitivity}). Moving the 15~m~s$^{-1}$ window by $\pm2.5$~m~s$^{-1}$ changes the five-level accuracy to 60.00\% or 58.82\% and the three-class accuracy to 82.35\% or 83.53\%; the three-class macro-F1 remains between 85.21\% and 86.23\%. Thus the coarse soft/medium/stiff discrimination is not an artifact of the exact 90--75~m~s$^{-1}$ endpoints, although fine five-level discrimination remains weak.

\begin{table}[!htbp]
\centering
\caption{Sensitivity of within-bed LOSO results to adjacent equal-width feature windows.}
\label{tab:window_sensitivity}
\resizebox{\linewidth}{!}{%
\begin{tabular}{rrrrr}
\toprule
Window (m s$^{-1}$) & Five-level acc. (\%) & Five-level macro-F1 (\%) & Three-class acc. (\%) & Three-class macro-F1 (\%) \\
\midrule
92.5--77.5 & 60.00 & 59.31 & 82.35 & 85.21 \\
90.0--75.0 & 54.12 & 52.95 & 83.53 & 86.23 \\
87.5--72.5 & 58.82 & 57.48 & 83.53 & 85.31 \\
\bottomrule
\end{tabular}%
}
\end{table}

The endpoint-sensitivity diagnostics reinforce the resolution limit rather than remove it. When the large-deformation $5\times10^6$~Pa endpoint is excluded, four-level LOSO correctly classifies 30/68 responses (44.12\% accuracy and balanced accuracy; 42.75\% macro-F1). When those same 68 responses are instead grouped into medium versus stiff classes, 54/68 are correctly classified (79.41\% for accuracy and balanced accuracy; 79.34\% macro-F1). Table~\ref{tab:endpoint_sensitivity} presents the four analyses together. The near-perfect separation of the soft endpoint raises the three-class score, but the medium--stiff result shows that coarse discrimination does not depend exclusively on it. Conversely, the low four-level score confirms that the three-class performance cannot justify fine modulus resolution.

\begin{table}[!htbp]
\centering
\caption{Within-bed LOSO classification and endpoint-sensitivity diagnostics. The two endpoint-excluded analyses omit $E=5\times10^6$~Pa.}
\label{tab:endpoint_sensitivity}
\resizebox{\linewidth}{!}{%
\begin{tabular}{lrrrr}
\toprule
Closed-set task & Correct/total & Accuracy (\%) & Balanced accuracy (\%) & Macro-F1 (\%) \\
\midrule
Five discrete levels & 46/85 & 54.12 & 54.12 & 52.95 \\
Four levels, endpoint excluded & 30/68 & 44.12 & 44.12 & 42.75 \\
Three broad classes & 71/85 & 83.53 & 86.27 & 86.23 \\
Two classes, endpoint excluded & 54/68 & 79.41 & 79.41 & 79.34 \\
\bottomrule
\end{tabular}
}
\end{table}

Because neighboring coordinates can share particles and surrounding structure, these metrics characterize transfer to an unseen coordinate within the present bed; they are not independent-bed generalization accuracies. The class-grouped results support coarse narrowing of the admissible Young's-modulus range under the present model conditions, whereas interpolation to unseen continuous $E$ values remains untested.

\FloatBarrier

The above results demonstrate robustness to \emph{local configurational variability within one settled bed realization}. They do not constitute a statistical test over independent global bed realizations, irregular particle shapes, or different particle-size distributions; those extensions require a broader calibration database.

\subsection{Mesoscale load-transmission mechanism of the Young's-modulus-sensitive response}
\label{sec:mesoscale}

The macroscopic response establishes a systematic Young's-modulus signature, but it does not by itself explain why the probe decelerates more strongly as $E$ increases. The mesoscale analysis defined in Section~\ref{sec:meso_methods} is therefore applied to the particle--particle contact population to examine high-load contacts, interface load sharing, strong-contact network morphology, and continuous internal load-transmission paths. These quantities are numerical explanatory variables rather than required onboard observables.

The numerical fraction of strong contacts is ROI dependent, whereas the conclusion of strongly heterogeneous load bearing is robust. In the near-field $2d\times4d$ region, only 1.48--3.72\% of the compressive particle--particle contacts are classified as strong, yet they carry 98.96--99.44\% of the local normal-force sum. In the $3d\times6d$ region the strong-contact fraction is 0.84--1.54\%, with 98.56--99.79\% of the local normal-force sum; in the $4d\times8d$ region the fraction falls to approximately 0.43--0.83\%, while the load fraction remains approximately 97.6--99.9\%. The robust result is therefore that the contact-force distribution is highly skewed and a small high-force subset dominates the modeled local load.

At the probe interface, the mean number of directly contacting particles decreases overall from 12.83 to 3.86 as $E$ increases, while $N_{\mathrm{eff}}$ decreases from 5.65 to approximately 2.5--2.9. The top-three load fraction rises from 63.0\% to above 94.7\%, reaching 96.7\% at $E=3.5\times10^8$~Pa. A slight reversal occurs at the highest $E$, so the defensible statement is that increasing the Young's-modulus-dependent contact response \emph{generally} localizes the interface load onto fewer dominant particles rather than producing a strictly monotonic localization metric.

\begin{figure}[H]
  \centering
  \includegraphics[width=0.98\linewidth]{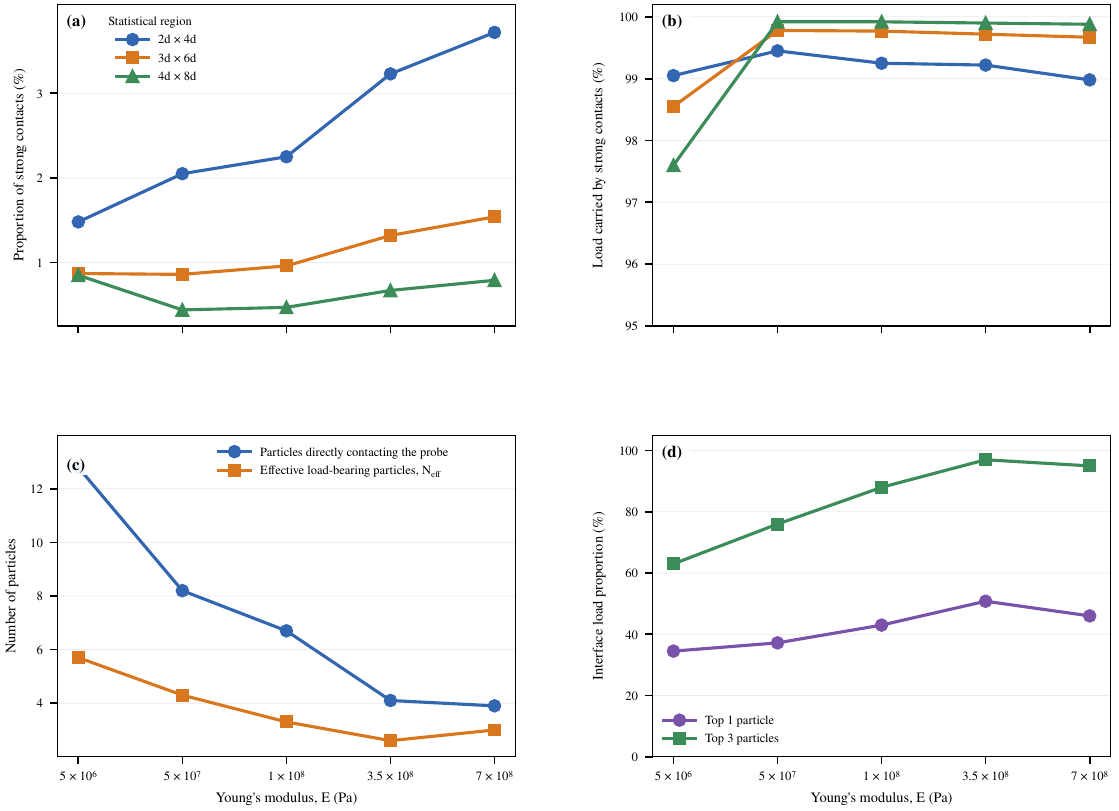}
  \caption{Strong-contact load heterogeneity and probe-interface load localization: (a) strong-contact fraction in three local ROIs, (b) fraction of local normal-force sum carried by strong contacts, (c) directly contacting and effective load-bearing particle counts, and (d) interface axial-load fraction carried by the largest one and largest three contributors.}
  \label{fig:interface}
\end{figure}

The graph-based morphology descriptors show that increasing $E$ produces more high-load contacts but does not simply generate one progressively longer vertical chain. Over the 16 common speed states, the median number of strong-contact edges increases from 3.5 to 10.0, while the number of connected components increases from 2.0 and remains near 5 for the higher-$E$ cases. The median axial extent of the probe-related strong-contact structure increases from about 0.088 to 0.151~m, whereas the radial extent is about 0.116--0.127~m for the first four $E$ levels and increases to about 0.184~m at the highest $E$.

The median axial-to-radial extent ratio remains between about 0.80 and 1.05 and is non-monotonic with $E$. The force-weighted axial descriptor likewise varies non-monotonically between approximately 0.50 and 0.71. The strong-contact network is therefore better described as a short-range three-dimensional structure with multiple load-bearing components and comparable axial and radial scales, rather than as a progressively longer and more vertical force chain as $E$ increases.

\begin{landscape}
\begin{figure}[p]
  \centering
  \includegraphics[width=0.98\linewidth]{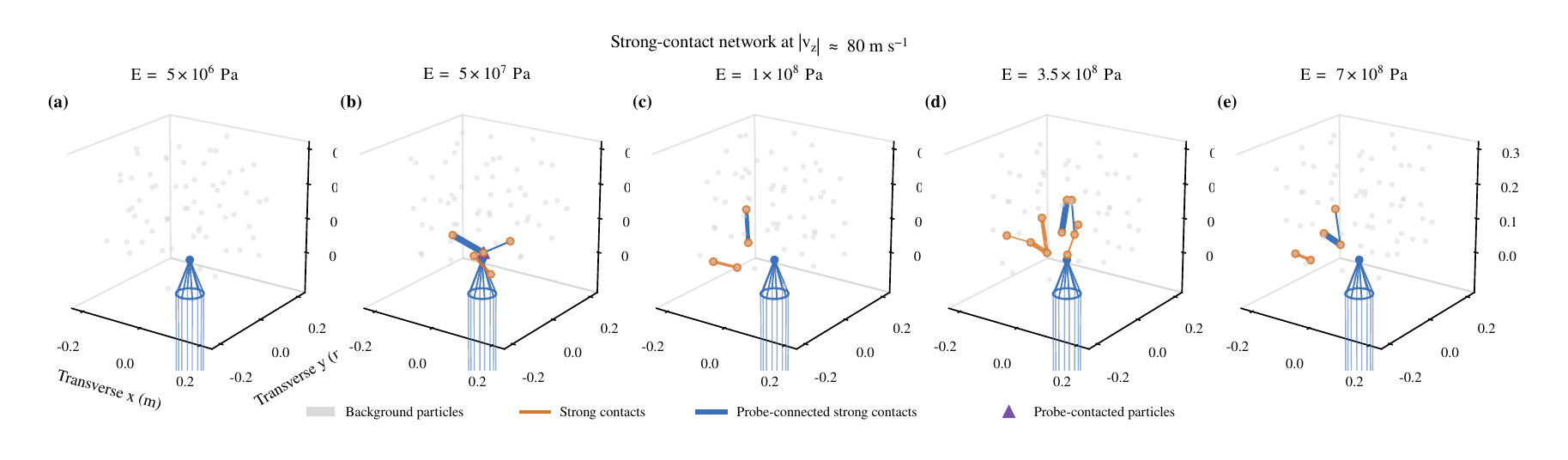}
  \vspace{2mm}
  \includegraphics[width=0.94\linewidth]{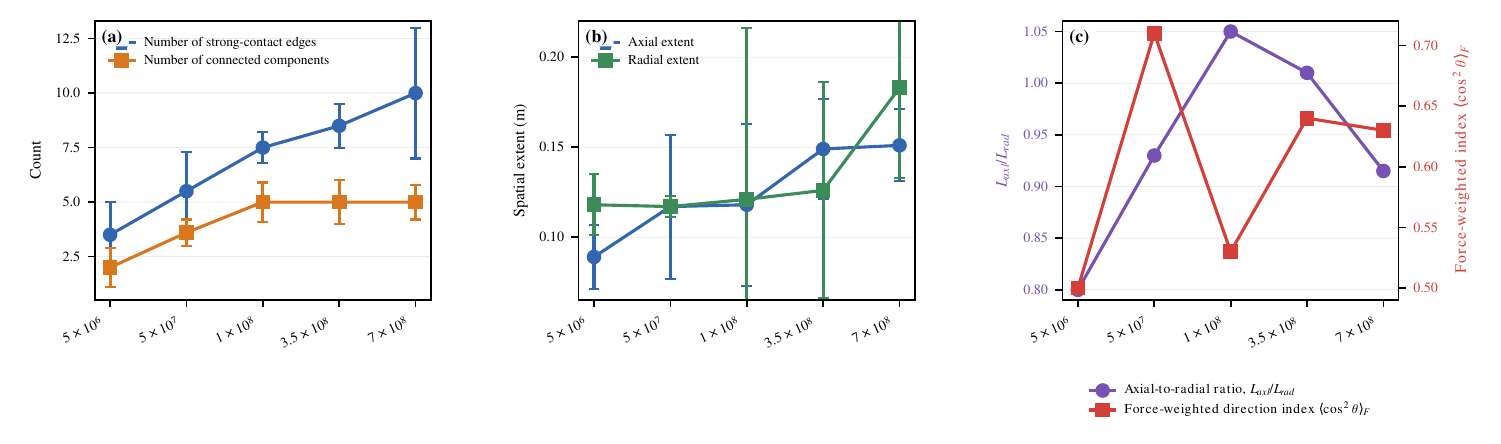}
  \caption{Strong-contact network morphology and statistics. The upper panel shows representative three-dimensional networks at $|v_z|\approx80$~m~s$^{-1}$; the lower panel summarizes strong-contact edge count, connected-component count, spatial extents, axial-to-radial ratio, and force-weighted axial direction over common speed states.}
  \label{fig:strong_network}
\end{figure}
\end{landscape}
\FloatBarrier

\begin{landscape}
\begin{figure}[p]
  \centering
  \includegraphics[width=0.98\linewidth]{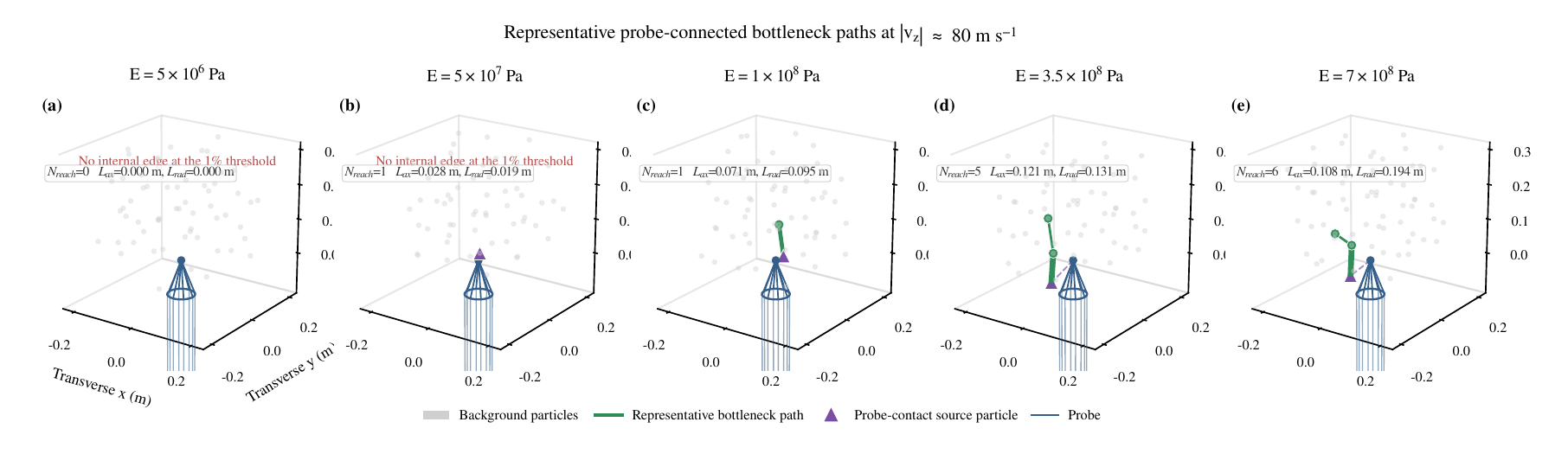}
  \caption{Representative 1\% widest paths at $|v_z|\approx80$~m~s$^{-1}$ for the five Young's-modulus cases. Green nodes and edges show representative keyframe bottleneck paths reconstructed from the weighted contact network; the annotations report keyframe path reachability and geometric extents. Full 90--75~m~s$^{-1}$ time-series statistics are reported in Fig.~\ref{fig:transmission_metrics} and Table~\ref{tab:meso}.}
  \label{fig:bottleneck3d}
\end{figure}
\end{landscape}

The widest-path analysis complements the strong-contact statistics by providing a graph-based connectivity descriptor: it asks how far a continuous high-capacity path can maintain a finite fraction of the instantaneous probe resistance in the weighted contact network. It should not be interpreted as an independent force-balance solution. Figure~\ref{fig:bottleneck3d} shows representative 1\% widest paths near $|v_z|\approx80$~m~s$^{-1}$. These panels are individual keyframes; the full 90--75~m~s$^{-1}$ time-series statistics are evaluated separately.

\FloatBarrier

\begin{figure}[H]
  \centering
  \includegraphics[width=0.90\linewidth]{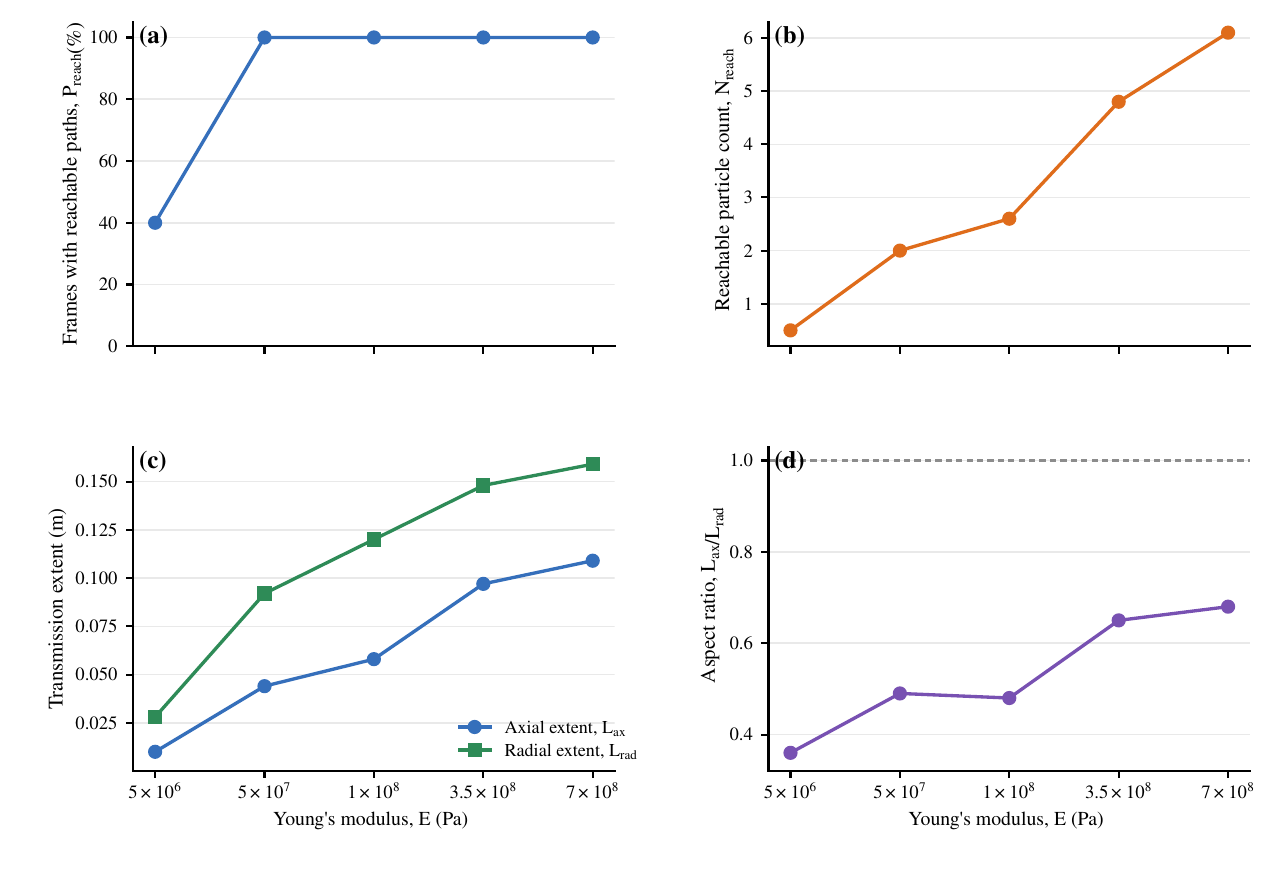}
  \caption{Graph-based load-transmission metrics at the representative 1\% instantaneous-probe-resistance threshold: (a) fraction of frames with at least one reachable node, (b) mean reachable-particle count, (c) axial and radial transmission extents, and (d) ratio of all-frame mean extents $\overline{L}_{\mathrm{ax}}/\overline{L}_{\mathrm{rad}}$. Frames without a reachable node contribute zero to both averaged extents; panel (d) does not average frame-wise ratios.}
  \label{fig:transmission_metrics}
\end{figure}

The full time-series statistics in Fig.~\ref{fig:transmission_metrics} show a systematic expansion of the graph-based transmission reach with increasing $E$. At the 1\% threshold, 40.4\% of the analyzed frames in the softest case contain at least one reachable node under the virtual-root definition, whereas all four higher-$E$ cases reach 100\%. When frames without a qualifying path contribute zero to the extent averages, $N_{\mathrm{reach}}$ increases monotonically from 0.50 to 6.06, $\overline{L}_{\mathrm{ax}}$ from 0.0096 to 0.108~m, and $\overline{L}_{\mathrm{rad}}$ from 0.027 to 0.160~m. All cases have $\overline{L}_{\mathrm{rad}}>\overline{L}_{\mathrm{ax}}$, demonstrating substantial radial spreading. Their ratio $\overline{L}_{\mathrm{ax}}/\overline{L}_{\mathrm{rad}}$ increases overall from about 0.36 to 0.67, indicating that transmission into the bed grows faster axially in a relative sense, while remaining strongly three-dimensional.

The same ordering of $N_{\mathrm{reach}}$, $L_{\mathrm{ax}}$, and $L_{\mathrm{rad}}$ with $E$ is preserved when the reachability threshold is varied over 0.01\%, 0.05\%, 0.1\%, 0.5\%, 1\%, 2\%, and 5\%. Increasing the threshold reduces the number of reachable particles and the spatial extents, as expected, but does not change the primary Young's-modulus ordering. The expansion is therefore not an artifact of choosing exactly 1\%.

\begin{table}[!htbp]
\centering
\caption{Representative mesoscale metrics for the five Young's-modulus cases. Strong-contact fraction and load fraction use the $R=2d$, $L=4d$ region. Transmission metrics use the 1\% instantaneous-probe-resistance threshold, with zero contribution from frames having no qualifying path.}
\label{tab:meso}
\resizebox{\linewidth}{!}{%
\begin{tabular}{rrrrrrrrr}
\toprule
$E$ (Pa) & $f_{sc}$ (\%) & $\eta_{sc}$ (\%) & $N_{probe}$ & $\eta_3$ (\%) & $P_{reach}$ (\%) & $N_{reach}$ & $L_{ax}$ (m) & $L_{rad}$ (m) \\
\hline
$5\times10^6$ & 1.48 & 99.06 & 12.83 & 63.0 & 40.4 & 0.50 & 0.010 & 0.027 \\
$5\times10^7$ & 2.04 & 99.44 & 8.16 & 75.6 & 100.0 & 1.98 & 0.044 & 0.091 \\
$1\times10^8$ & 2.25 & 99.23 & 6.65 & 87.0 & 100.0 & 2.57 & 0.057 & 0.120 \\
$3.5\times10^8$ & 3.23 & 99.19 & 4.13 & 96.7 & 100.0 & 4.85 & 0.096 & 0.147 \\
$7\times10^8$ & 3.72 & 98.96 & 3.86 & 94.7 & 100.0 & 6.06 & 0.108 & 0.160 \\
\bottomrule
\end{tabular}%
}
\end{table}

To determine whether the mechanism is unique to one penetration location, three separately retained mechanism cases with low, intermediate, and high deceleration are compared at $E=3.5\times10^8$~Pa. Their nominal bed-plane coordinates are $(0.1,0.1)$, $(0,0)$, and $(-0.1,-0.1)$~m for P02\_1, P01\_0, and P03\_2, respectively; they are used as illustrative response cases rather than as a statistical subsample of the 85-case classifier. Using the same 90--75~m~s$^{-1}$ fitting procedure, the characteristic decelerations are 14607, 15960, and 21560~m~s$^{-2}$, a 47.6\% difference between the high- and low-response positions. Meanwhile, the mean number of directly contacting particles decreases from 5.53 to 3.31 and the top-three interface load fraction rises from 86.9\% to 99.5\%. Thus the common tendency toward sparse dominant interface loading persists even though the macroscopic responses differ.

The stronger location dependence appears in the spatial realization of internal transmission. At the 1\% threshold, $N_{\mathrm{reach}}$ increases from 1.70 for the low-response position to 4.85 and 6.39 for the intermediate- and high-response positions. The radial extent increases from 0.070 to 0.202~m, while the axial extents are 0.046, 0.096, and 0.090~m and therefore do not increase strictly with response magnitude. The force-weighted axial descriptor decreases from about 0.747 to 0.384. A larger response is therefore associated with more reachable particles and broader radial/oblique transmission, not necessarily with a longer or more vertically aligned chain.

\begin{figure}[!htbp]
  \centering
  \includegraphics[width=0.90\linewidth]{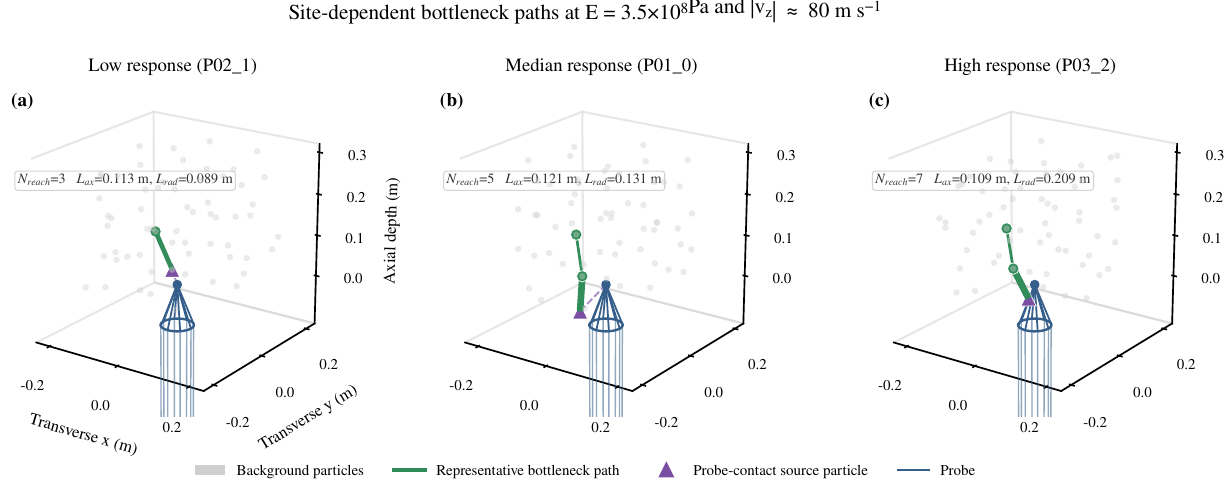}
  \caption{Representative 1\% widest paths at $|v_z|\approx80$~m~s$^{-1}$ for three response locations at $E=3.5\times10^8$~Pa: (a) low response P02\_1, (b) intermediate response P01\_0, and (c) high response P03\_2. The keyframe paths illustrate different spatial realizations of load transmission at the same $E$.}
  \label{fig:three_sites}
\end{figure}

\begin{table}[!htbp]
\centering
\caption{Macroscopic response and transmission metrics for three representative locations at $E=3.5\times10^8$~Pa.}
\label{tab:three_sites}
\resizebox{\linewidth}{!}{%
\begin{tabular}{lrrrrrrrr}
\toprule
Response & Case & $a$ (m s$^{-2}$) & $N_{probe}$ & $\eta_3$ (\%) & $N_{reach}$ & $L_{ax}$ (m) & $L_{rad}$ (m) & $A_F$ \\
\hline
Low & P02\_1 & 14607 & 5.53 & 86.9 & 1.70 & 0.046 & 0.070 & 0.747 \\
Intermediate & P01\_0 & 15960 & 4.13 & 96.7 & 4.85 & 0.096 & 0.147 & 0.688 \\
High & P03\_2 & 21560 & 3.31 & 99.5 & 6.39 & 0.090 & 0.202 & 0.384 \\
\bottomrule
\end{tabular}%
}
\end{table}

Under the adopted Hertz-type nonlinear normal-loading relation, increasing $E$ increases elastic contact stiffness. The computed contact-force field simultaneously shows stronger dominant loading, a general tendency toward more localized probe-interface load sharing, and a larger graph-based reach of high-capacity connected paths. These changes are consistent with the increase of $a$ with $E$ and provide a structural interpretation of the modeled response; the network descriptors are not, by themselves, an independent causal proof. Different penetration locations alter the exact contact connectivity and spatial realization of these high-capacity paths, superimposing configuration-dependent scatter on the common modulus trend.

A post-processing validity check was also performed on the retained contact data over the 90--75~m~s$^{-1}$ interval. Using the physical particle geometry, the 90th-percentile probe--particle overlap normalized by particle diameter was approximately 0.195, 0.095, 0.093, 0.053, and 0.041 from the lowest to the highest $E$ level. These are 90th-percentile physical geometric overlaps, conditional on positive probe--particle overlap, and refer to the 0.10-m particle diameter; they are not engineering strains of an individual rock. At 100~m~s$^{-1}$ initial probe speed, the chosen compliant contact laws accommodate the rapid impact partly through large numerical interpenetration. The $5\times10^6$~Pa case is therefore treated as a pronounced large-deformation numerical end member; the 0.095 and 0.093 diameter-normalized values in the next two groups also call for caution in interpreting Hertzian contact deformation quantitatively. Even the stiffer cases are not independently validated against dynamic clast-contact or fragmentation experiments. Aggregate force--overlap fits for both particle--particle and probe--particle contacts have log--log slopes between 1.492 and 1.501 with $R^2>0.9998$, confirming that the implementation follows the prescribed $n=1.5$ force--overlap form. This implementation check does not establish physical validity of the Hertz small-deformation approximation at the largest overlaps. As a separate boundary-localization check, the 1\% high-capacity probe-connected paths reconstructed in the same early-response window remained within approximately 0.24~m axially and 0.23~m radially from the probe over the five representative $E$ cases, far inside the $2\times2\times4$~m domain. The dominant load-transmission structures used in the mechanism analysis therefore do not approach the container boundaries, although this localization check is not a substitute for a full domain-size sweep.

This interpretation remains bounded by the contact model. Large soft-sphere overlaps in the lowest-$E$ cases represent large local compliance in the soft-sphere representation and should not be interpreted as literal elastic compression of intact decimeter-scale rocks. Likewise, the dust-bridge adhesion representation $F_{\mathrm{adh}}=cA_{\mathrm{eff}}$ is a coarse-grained contact-level approximation and is not a microscopic van der Waals force--separation law. The network quantities therefore explain Young's-modulus ordering \emph{under the present modeled conditions}; quantitative transfer to irregular, polydisperse, or fragmenting asteroid regolith requires further calibration.

\subsection{Summary of the results}
\label{sec:results_summary}

In the exploratory single-reference-configuration screen at 100~m~s$^{-1}$, particle Young's modulus produces the clearest early velocity-decay signature among the six tested parameter families; rolling friction, like the other non-$E$ parameters, does not change this ranking. In the 85-case ensemble formed by five $E$ levels and 17 local penetration configurations within one parent bed, the fitted 90--75~m~s$^{-1}$ characteristic deceleration increases systematically with $E$, while local configuration produces secondary but non-negligible scatter. The location-averaged response is well represented by an empirical power-law relation over the investigated range, and the observed blocked sum-of-squares partition shows that the Young's-modulus term is the largest modeled component within the present paired ensemble. The five-level and three-class cross-validations quantify the resolution supported by the current sparse calibration: the three broad modulus classes provide a substantially more reliable coarse narrowing of the admissible $E$ range, whereas the five-level results show pronounced overlap at higher stiffness. Finer inference would require both more representative calibration and evidence that the overlapping response distributions can in fact be separated; denser sampling alone does not guarantee this.

At the particle scale, increasing $E$ strengthens dominant contact loading, generally localizes interface resistance onto fewer particles, and expands the population and spatial reach of continuous contact paths capable of carrying a finite fraction of the instantaneous probe load. The exact branch directions remain controlled by local granular configuration and do not show a simple monotonic rotation toward the probe axis. Together, these results establish a numerical evidence chain from an onboard-measurable early deceleration to coarse statistical constraint of Young's modulus under the present controlled conditions, followed by a particle-scale structural interpretation of the observed trend.

\section{Implications for in situ regolith characterization}
\label{sec:implications}

The results show that the early probe response contains systematic information on the model particle Young's modulus $E$, while local particle configuration introduces appreciable scatter. The practical question is how the probe's own motion can be used to constrain $E$ without requiring direct observation of the mesoscale contact network.

\subsection{In-situ inference of Young's modulus}
\label{sec:mission_inference}

Spacecraft--surface interactions have already shown that vehicle dynamics can constrain the mechanical state of small-body near-surface material. The interaction of OSIRIS-REx with Bennu, for example, provided evidence for an extremely loose near subsurface with near-zero cohesion \citep{Walsh2022}, and kinetic penetration has been considered as a route to in situ regolith-property characterization \citep{Jiang2022}. Unlike quasi-static probing, a kinetic probe acquires its energy before impact and may reduce the need for an apparatus that sustains a large reaction force against a microgravity surface. Whether this yields a simpler or lighter flight system remains to be established through mission-specific engineering studies.

Practical use of the method can be separated into two stages: pre-mission calibration and in situ constraint of the admissible modulus range. Before flight, numerical simulations and dedicated experiments can be used to construct a response database over representative Young's-modulus values while varying local particle arrangements and other regolith conditions expected to affect the signal. For each part of the calibrated $E$ domain, the distribution of the early probe response can then be quantified. The 85-case ensemble provides an initial within-bed demonstration of this statistical calibration concept rather than a complete mission database.

For a future penetration experiment, the proposed response-based estimator would require a calibrated early axial velocity or acceleration record. The characteristic deceleration $a$ is extracted from the calibrated early-response window and compared with the pre-mission database to constrain the supported particle Young's-modulus range. Depending on calibration coverage and the degree of overlap among response distributions, the output may be expressed as a best-supported calibration level, a plausible interval, or a probability distribution. Using the mean responses of the five $E$ cases, the 90--75~m~s$^{-1}$ interval corresponds approximately to 0.67--4.48~ms, so the informative signal is concentrated in the first few milliseconds of sustained contact. The high $R^2$ of the mean response fit should not be interpreted as equivalent to high precision for an individual penetration. Coordination, strong-contact structure, and widest-path measures remain explanatory quantities used in simulation; they are not required as in situ measurements for the proposed response-based comparison.

The five-level, four-level endpoint-excluded, and class-grouped results in Section~\ref{sec:robustness} illustrate different resolutions of the same particle Young's-modulus range-constraint problem. Here, range constraint means narrowing the candidate set of calibrated Young's-modulus levels to a broader class. The reported class accuracy applies to the sampled levels; it does not establish coverage or identification accuracy for unsampled values within or between the classes. This coarse inference retains the same Young's-modulus characterization objective. The 79.41\% medium-versus-stiff accuracy after excluding the soft endpoint shows that the coarse signal is not solely an endpoint artifact, whereas the corresponding 44.12\% four-level accuracy confirms substantial overlap among adjacent higher-modulus levels. For the empirical mean relation $\bar a\propto E^{0.388}$, first-order error propagation gives $\Delta E/E\approx(1/0.388)\,\Delta a/a\approx2.58\,\Delta a/a$; response uncertainty is therefore amplified when the mean relation is inverted. A denser and more representative calibration database can characterize the response distributions more faithfully and may improve resolution, but it does not by itself guarantee separation of intrinsically overlapping distributions. Resolution improvement should therefore rely on pre-mission numerical and experimental calibration together with explicit uncertainty quantification, rather than on large numbers of repeated penetrations after arrival.

The 17 simulated locations primarily characterize how local particle arrangement broadens the response distribution within one deliberately controlled parent-bed realization. Reusing the same bed is advantageous for paired comparison because it suppresses unrelated between-bed variability, but these cases do not establish population-level uncertainty across independent regolith realizations. A future independent-bed check would regenerate the packing while retaining the container, nominal probe coordinates, and comparison protocol; moving the present sites toward the walls is not an equivalent check. In a broader pre-mission calibration, additional independent beds, particle shapes, gradations, and packing states should be included so that one or a small number of in situ penetration records can be compared with response distributions that already account for modeled structural variability.

\subsection{Applicability and model limitations}
\label{sec:limitations}

The inferred Young's modulus is the material parameter $E$ assigned to the rubble particles in the DEM model. Under the Hertz-type nonlinear normal-loading relation adopted here, $E$ enters the contact-pair effective modulus $E^*$ and directly controls the elastic contact response. Quantitative inference must nevertheless be interpreted together with the modeling assumptions used to construct the calibration database. Particle shape, size distribution, internal porosity, fragmentation, packing state, unresolved dust, local contact configuration, gravity level, and probe conditions can all change the measured response and hence the calibration between $a$ and $E$.

The model uses monodisperse spherical particles and does not include particle fragmentation, clast plasticity, or shape evolution. The overlap diagnostics in Section~\ref{sec:mesoscale} identify the $5\times10^6$~Pa case as a pronounced large-deformation numerical end member; substantial probe--particle overlap is also found at $5\times10^7$ and $1\times10^8$~Pa. At this impact speed, the inferred $E$ is thus a contact-law parameter calibrated to the present coarse-grained dynamics, not an independently validated measurement of intact-clast elastic modulus. The absence of fragmentation is a consequential approximation when deformation and local loading are large. Overlap alone does not demonstrate that a particular rock would fracture, but this model cannot determine whether or how it would do so. Fracture could redistribute probe resistance and alter the $a$--$E$ calibration and its class separation; the present inference is therefore conditional on the non-fragmenting particle assumption. Introducing breakage would require a separately calibrated strength criterion and fragment/contact representation, rather than a reinterpretation of the current 85 intact-particle runs. A quantitative claim about real surface-rock modulus requires independent dynamic contact, penetration, or fragmentation calibration over the relevant scales.

The dust-bridge representation $F_{\mathrm{adh}}=cA_{\mathrm{eff}}$ is physically motivated at the coarse-grained level by fine-regolith bridges but is not a microscopic van der Waals force--separation law. The fixed area factor $\beta=1$ and the scanned $c$ values therefore require material- and particle-system-specific calibration before quantitative transfer to a particular asteroid. 

Rolling friction is included with the other material/contact parameters in the OFAT screen at one baseline configuration. The penetration time-step check over $\Delta t=1\times10^{-7}$--$4\times10^{-7}$~s supports the adopted production step but is not a numerical-convergence proof over the full parameter space. No formal sensitivity sweeps were performed for parameter interactions, coupling-transfer settings, impact velocity, or domain size; feature-window placement was checked only over the two adjacent equal-width windows reported in Section~\ref{sec:robustness}. The observed high-capacity paths remain well inside the container, which supports limited direct boundary participation during the analyzed interval but is not a substitute for a domain-size sweep. The $a$--$E$ relation is consequently a calibration for the present probe, particle-bed, contact-law, numerical, and environmental conditions rather than a universal constitutive law.

The inference analysis also assumes ideal access to the simulated axial velocity history. Instrument bandwidth, accelerometer or velocity-estimator noise, sampling and synchronization error, quantization, filtering, and onboard feature-extraction latency are not modeled. Accordingly, the LOSO metrics quantify numerical separability of the simulated response, not end-to-end flight-sensor performance. These effects should be introduced explicitly in future hardware-in-the-loop or penetration experiments.

\subsection{Need for direct experimental validation}
\label{sec:validation_need}

Direct experimental validation of the proposed early-response--Young's-modulus relation is an important next step. Ground-based high-speed penetration tests can first close the loop between probe structure, instrumentation, feature extraction, and the numerical model. Reduced-gravity tests can then constrain the influence of gravity on the calibration. The detailed apparatus, launcher, sensor package, and scaling strategy should be designed around a specific mission and are outside the scope of the present study. The contribution here is to show numerically, within the adopted DEM--MBD framework, that the early probe response retains useful coarse-grained information on the model particle Young's modulus despite within-bed configurational variability, and to identify particle-scale structural changes that accompany that signal and its uncertainty. Direct high-speed experiments and independent-bed validation are still required before mission-level quantitative inversion can be claimed.

\section{Conclusions}
\label{sec:conclusions}

This study evaluated whether the early axial response of a kinetic penetrator can constrain asteroid-regolith particle stiffness, parameterized through the prescribed particle Young's modulus $E$ in the present DEM model, in the presence of local granular variability. The principal conclusions are as follows.

\begin{enumerate}
\item In an exploratory single-reference-configuration OFAT screen of particle Young's modulus, Poisson's ratio, particle--particle restitution, static friction, rolling friction, and adhesion strength, the Young's-modulus cases show the clearest and most systematic separation in the early probe velocity response at 100~m~s$^{-1}$. The subsequent paired 17-location analysis confirms that the characteristic deceleration $a$ extracted from the common 90--75~m~s$^{-1}$ interval increases systematically with $E$ within the controlled parent bed. The location-averaged response is described by an empirical calibration over the investigated range, not by a universal constitutive law.

\item Multi-position simulations show that local particle arrangement introduces appreciable scatter into individual penetration responses, while a systematic Young's-modulus trend remains within the present paired ensemble. Five-level within-bed LOSO gives 54.12\% accuracy (52.95\% macro-F1), whereas the exploratory three-class grouping gives 83.53\% accuracy (86.23\% macro-F1). After exclusion of the large-deformation soft endpoint, medium-versus-stiff accuracy remains 79.41\%, but four-level accuracy is only 44.12\%. The evidence therefore supports coarse closed-set narrowing of the Young's-modulus range more strongly than high-resolution or continuous inversion. Independent-bed transfer and interpolation to unseen $E$ values remain unvalidated.

\item Mesoscale analysis provides a particle-scale structural interpretation of the Young's-modulus-sensitive signal. Increasing $E$ generally concentrates probe-interface resistance onto fewer dominant particles and increases the graph-based reach of high-capacity connected paths. These network quantities describe changes in the computed contact-force organization; they are not an independent mechanical equilibrium solution or standalone causal proof. The detailed orientation and topology of these paths remain controlled by local granular configuration and do not evolve toward a single universal force-chain geometry.

\item The results motivate a mission concept combining pre-mission calibration with a limited number of in situ penetrator measurements. A representative numerical--experimental database spanning particle Young's modulus together with independent bed realizations and relevant variations in grading, shape, packing, gravity, impact condition, and probe configuration could allow one or a small number of early probe-response records to constrain the admissible Young's-modulus range. Direct high-speed experiments, sensor-chain validation, impact-speed/window sensitivity studies, and broader model validation remain necessary before quantitative mission-level inversion can be claimed.
\end{enumerate}

This study provides numerical evidence that the early motion of a kinetic penetrator retains a stiffness-dependent signature despite local granular variability. Its principal implication is therefore a calibrated and experimentally testable route toward in situ narrowing of the admissible particle Young's-modulus range of an asteroid-regolith layer. As a potential low-complexity precursor reconnaissance approach, kinetic penetration could provide subsurface response information before deeper-contact exploration of poorly characterized bodies. Mission safety, cost, and quantitative performance nevertheless require dedicated experimental and engineering assessment.

\section*{Author contributions}
\textbf{Zichen Wang:} Research methodology, numerical investigation, data analysis, visualization, and preparation of the original manuscript.
\textbf{Yang Yu:} Overall research supervision and guidance throughout the study, including the development of the research approach and methodology.
\textbf{Chenyang Huang and Zhijun Song:} Guidance on the research direction and conceptual development of the study.
\textbf{Yutian Wu:} Academic consultation and assistance in resolving research-related questions.
\textbf{Xiaojing Zhang and Xuhui Zhang:} Contributions to the contextualization of the study through discussions of current developments and emerging research directions in the field.

\section*{Acknowledgements}
This work was supported by the National Natural Science Foundation of China (Grant No. 12672056).

\section*{Declaration of competing interest}
The authors declare that they have no known competing financial interests or personal relationships that could have appeared to influence the work reported in this paper.

\section*{Data availability}
The exported simulation data and processed datasets supporting the findings of this study are available from the corresponding author upon reasonable request.

\section*{Declaration of generative AI and AI-assisted technologies in the writing process}
During the preparation of this work, the authors used ChatGPT and Codex (OpenAI) to assist with English-language editing, improvement of readability, manuscript organization, consistency checking, and figure formatting. No AI tool was used to generate the numerical simulation outputs. All AI-assisted content and quantitative results were critically reviewed and verified by the authors, who take full responsibility for the scientific content, interpretation, and final version of the manuscript.

\end{document}